\documentclass[journal]{IEEEtran}
\usepackage{epsfig,epsf,epstopdf}
\usepackage{hyperref}

\usepackage{amsthm}

\newtheoremstyle{boldhead}
  {} % Space above
  {} % Space below
  {\normalfont} % Body font - non-italic
  {} % Indent amount
  {\bfseries\itshape} % Theorem head font - bold and italic
  {.} % Punctuation after theorem head
  { } % Space after theorem head
  {\thmname{#1}\thmnumber{ #2}\thmnote{ \normalfont(#3)}} % Theorem head spec (for definitions and remarks)

\theoremstyle{bolditalichead}
\newtheorem{theorem}{Theorem}
\newtheorem{problem}{Problem}
\newtheorem{assumption}{Assumption}
\newtheorem{definition}{Definition}
\newtheorem{remark}{Remark}
\newtheorem{lemma}{Lemma}

\graphicspath{ {./Figs/} }
\usepackage[cmex10]{amsmath}
\usepackage{amsfonts}
\usepackage{amssymb}

\usepackage[noend]{algpseudocode}
\makeatletter
\def\BState{\State\hskip-\ALG@thistlm}
\makeatother
\usepackage{cite}
\usepackage{paralist}
\usepackage{color}

\definecolor{blue}{rgb}{0, 0.1, 0.7}
\usepackage{mathrsfs}
\usepackage{multirow}
\usepackage{subfigure}
\usepackage{graphicx}
\usepackage{subcaption}
\usepackage{caption}
\usepackage[font=small,justification=justified,singlelinecheck=false]{caption}

\ifCLASSINFOpdf
 
\else
 
\fi

\begin{document}

\title{Cyber-physical Defense for Heterogeneous Multi-agent Systems Against Exponentially Unbounded Attacks on Signed Digraphs}

\author{Yichao Wang, Mohamadamin Rajabinezhad, Yi Zhang, and Shan Zuo% <-this % stops a space
\thanks{Yichao Wang, Mohamadamin Rajabinezhad, Yi Zhang and Shan Zuo are with the department of electrical and computer engineering, University of Connecticut, CT 06269, USA. (E-mails: yichao.wang@uconn.edu; mohamadamin.rajabinezhad@uconn.edu; yi.2.zhang@uconn.edu; shan.zuo@uconn.edu.)}
}

\maketitle
\begin{abstract}
Cyber-physical systems (CPSs) are subjected to attacks on both cyber and physical spaces. In reality, attackers could launch any time-varying signals. Existing literature generally addresses bounded attack signals and/or bounded-first-order-derivative attack signals. In contrast, this paper proposes a privacy-preserving fully-distributed attack-resilient bilayer defense framework to address the bipartite output containment problem for heterogeneous multi-agent systems (MASs) on signed digraphs, in the presence of exponentially unbounded false data injection (EU-FDI) attacks on both the cyber-physical layer (CPL) and observer layer (OL). First, we design attack-resilient dynamic compensators that utilize data communicated on the OL to estimate the convex combinations of the states and negative states of the leaders. To enhance the security of transmitted data, a privacy-preserving mechanism is incorporated into the observer design. The privacy-preserving attack-resilient observers address the EU-FDI attacks on the OL and guarantee the uniformly ultimately bounded (UUB) estimation of the leaders' states in the presence of the eavesdroppers. Then, by using the observers' states, fully-distributed attack-resilient controllers are designed on the CPL to further address the EU-FDI attacks on the actuators. The theoretical soundness of the proposed bilayer resilient defense framework is proved by Lyapunov stability analysis. Finally, a comparative case study for heterogeneous MASs and the application in DC microgrids as a specific case study validate the enhanced resilience of the proposed defense strategies.
\end{abstract}

% Define a new environment for Note to Practitioners to mimic the abstract
\newenvironment{NoteToPractitioners}
  {\renewcommand\abstractname{Note to Practitioners}% Change abstract title
   \begin{abstract}}
  {\end{abstract}}

\begin{IEEEkeywords}
Cyber-physical defense, heterogeneous multi-agent systems, resilient control, signed digraph, exponentially-unbounded attacks, privacy preserving.    
\end{IEEEkeywords}

\section{Introduction}
In recent decades, multi-agent systems (MASs) have seen substantial advancements and have become a key research area in the system and control community due to their promising applications, such as multi-robot systems, sensor networks, smart grids, microgrids, social networks, task migration of many-core microprocessors, coordination of the charging of electric vehicles, and distributed heating, ventilation, and air conditioning optimization \cite{shi2015cooperative,guo2022observer,xue2021distributed,huang2023formation,chen2019control}. For instance, distributed consensus of unmanned surface vehicles under heterogeneous unmanned aerial vehicle-unmanned surface vehicle multi-agent systems cooperative control scheme is studied in \cite{xue2021distributed}. And formation control for unmanned aerial vehicle-unmanned surface vessel heterogeneous system with collision avoidance performance is studied in \cite{huang2023formation}. The dynamics of interactions in MASs are crucial for understanding and optimizing system performance. Significant progress has been made in achieving consensus and other collective behaviors in MASs across various network types, such as fixed, time-varying, and leader-follower networks, as demonstrated by \cite{knorn2015overview}. Despite these advancements, cooperative control within MASs remains an area that deserves more in-depth exploration \cite{wang2024multi}. Understanding cooperative control is essential as it directly affects the efficiency and effectiveness of collaborative tasks in complex environments. 

In the systems of most of the studies, the interaction topology is typically represented by an unsigned graph, assuming that the interaction weights among the agents are positive. This representation, while effective in a broad sense, may not always encapsulate the complexities of certain real-world systems. Bridging this gap requires a deeper exploration of MASs in scenarios involving both cooperative and antagonistic interactions. For instance, in social networks or political opinion dynamics within two-party systems \cite{wasserman1994social}, individuals' ideas or views do not uniformly align. A similar scenario is observed in antagonistic robotic networks \cite{qin2016bipartite}, gene transcriptional regulation biological networks \cite{jayaraman2016blue}, and predator-prey interactions \cite{zhai2019survival}, where agents exhibit both cooperative and antagonistic behaviors. 

When considering multiple leaders and followers in heterogeneous MASs that communicate on signed digraphs with both cooperative and antagonistic interactions, the classical bipartite consensus problems are transformed into bipartite output containment problems \cite{gao2023distributed,zuo2018bipartite}. For example, a swarm of UAVs (unmanned aerial vehicles) with both cooperative and antagonistic interactions can be modeled by signed communication digraphs with both positive and negative edge weights. This consideration generalizes the containment control of MASs by incorporating signed communication graphs. Communication is one of the key elements in bipartite output containment problems. Since, in some cases, MASs are deployed in sparse communication networks, where distributed control is usually deployed, while limited connectivity among agents creates significant security vulnerabilities. In such environments, local agents lack a global perspective and rely heavily on partial and potentially compromised information from their neighbors. False data injection (FDI) attacks are one of the most prominent cyber threats in such settings and the prevalence of FDI attacks has increased with the growing reliance on distributed systems and internet of things networks, as attackers exploit the lack of centralized control and the inherent vulnerabilities in communication protocols, posing severe risks to the stability and performance of MASs \cite{teixeira2015secure}. These attacks manipulate system data and measurements, compromising data integrity and misleading controllers into making incorrect decisions. The impact of FDI attacks can be particularly devastating, as they can destabilize the system, and lead to catastrophic failures such as blackouts in power grids or compromised operations in automated vehicles \cite{yan2016power,chen2024secure}.

When a system is attacked, detection-based mechanisms are often deployed to identify malicious activities. After detection, the system typically has two options: either the compromised agents are isolated or removed from the system \cite{chen2019resilient}, or the signals are compensated for without isolation through control mechanisms. For critical systems where the removal or isolation of agents can compromise the system’s overall functionality and cohesiveness. However, isolating or removing agents can compromise the system's functionality, especially in critical infrastructures where the loss of even a single agent may disrupt operations. Alternatively, compensation-based methods aim to mitigate the impact of attack signals through control mechanisms.

Existing resilient control strategies, such as $H_\infty$ control and fault-tolerant control, are primarily designed to address bounded disturbances or attack signals. These methods, however, cannot fully compensate for unbounded signals and often fail to prevent system instability or failure. Furthermore, even some approaches consider unbounded signals. such as the ones in \cite{zuo2020resilient,zuo2021resilient,zuo2023resilient}, but they frequently assume that the first-order time derivative of the attack signal is bounded, limiting their applicability to fully compensate the attack signals. Because, in reality, adversaries can inject \textit{any time-varying signal} into systems via software, CPU, DSP, or similar platforms. The signals could be \textit{unbounded}. For instance, in \cite{chen2021bipartite}, it is investigated that bipartite containment control in networked agents under denial-of-service attacks, employing dynamic signed digraphs to model variable communication links. In \cite{wu2023bipartite}, it is addressed that bipartite containment control in nonlinear MASs with time-delayed states under impulsive FDI attacks, and with Markovian variations in communication topology. 
In  \cite{jiang2021fully}, it is studied that dual-terminal dynamic event-triggered bipartite output containment control in heterogeneous linear MASs with actuator faults. The literature \cite{wang2022novel} introduces an innovative adaptive bipartite consensus tracking strategy for MASs under sensor deception attacks. In \cite{zhao2023self}, it is  explored that the design of bipartite formation containment tracking in heterogeneous MASs, considering external disturbances and inaccessible state vectors. In \cite{cheng2023adaptive}, it is investigated that adaptive bipartite output containment in heterogeneous MASs through a signed graph and a protocol with a distributed observer, addressing unmeasurable yet bounded inputs in leader dynamics. 

Moreover, existing detection-based methods often rely on restrictive assumptions, such as limiting the number of attacked agents. These constraints limit their applicability, particularly in scenarios where the adversary compromises a significant portion—or even all—of the network. Besides, an observer design is generally needed to address the output regulation problem for heterogeneous MASs by estimating the leaders' states. However, existing literature on heterogeneous MASs typically assumes that the observers remain intact against cyber-attacks, which is not practical. Although a few studies \cite{yang2023distributed} consider attacks on the observer, they typically assume that the first time derivative of the attack signals is bounded. This assumption restricts the applicability of the proposed countermeasures in more general scenarios.

Besides the defense capability against attack, privacy is another key component in MASs. For instance, in sensitive applications such as battlefield scenarios, some sensitive information, such as initial states of leaders and trajectories of
is often intended to remain confidential from other agents and outside world. Hence, preserving the data privacy of the leader vehicles is crucial \cite{chong2019tutorial}. Several approaches have been proposed in recent years to address this issue. One approach is based on cryptography, where encrypted messages are exchanged among agents using methods such as trusted third parties \cite{lazzeretti2014secure}, obfuscation \cite{ambrosin2017odin}, or distributed cryptography schemes \cite{ruan2017secure}. Another approach relies on differential privacy \cite{dwork2006differential,dwork2014algorithmic}, which involves adding noise from an appropriate source to the state transmitted by an agent. This ensures that even if the value is publicly broadcasted, the knowledge an observing agent can acquire about the true state is limited to a predetermined precision.
This method has been extensively studied in the context of the average consensus problem \cite{cortes2016differential,gupta2017privacy,huang2012differentially,nozari2017differentially,altafini2019dynamical}.

% For instance, in \cite{manitara2013privacy,mo2016privacy,rezazadeh2018privacy}, encryption is achieved through perturbations with zero sum (or integral) over time. A third approach focuses on understanding what is observable and what is not at a node \cite{alaeddini2017adaptive,pequito2014design}, aiming to guarantee privacy through a loss of observability.

This paper addresses the bipartite output containment problem for heterogeneous MASs under the exponentially unbounded false data injection (EU-FDI) attacks, incorporating the critical yet often neglected aspect of privacy preservation. A bilayer defense architecture is proposed, comprising a CPL and an OL, to enhance system resilience against EU-FDI attacks. Unlike existing studies, this work ensures the preservation of privacy by safeguarding the leaders' states, the convex combinations of the leaders' states which incorporate the graph topology information, from disclosure, thereby providing a comprehensive solution that balances robustness against EU-FDI attacks and stringent privacy requirements.

The main contributions of this paper are fourfold:

\begin{itemize}
\item A general privacy-preserving attack-resilient bipartite output containment (PABOC) problem is first formulated, considering both cooperative and antagonistic interactions among agents, removing the assumption that the edge weights have the same sign. To the best of the authors' knowledge, the rigorous mathematical proof is provided {\textit{for the first time}}, which asserts that the PABOC problem is solved by ensuring that the neighborhood bipartite output containment error is uniformly ultimately bounded (UUB).

\item This work introduces a privacy-preserving mechanism in the OL design, applying adaptive masking functions to the transmitted data to ensure confidentiality during communication on the digital OL, which is vulnerable to eavesdroppers. By employing time-varying adaptively tuned parameters in the mask function for data transmission among followers, the proposed privacy-preserving mechanism dynamically enhances privacy preservation, making it more difficult for eavesdroppers to infer critical system information from intercepted data. This design is particularly suited for applications, such as UAV swarms, where safeguarding vehicles' initial locations and trajectories is crucial for mission integrity in adversarial environments.

% {\color{red}\item Privacy-preserving mechanisms are integrated into the bipartite output containment framework. A novel output mask is proposed to protect agents' initial states, which avoids the use of random noise and ensures that all agents converge exactly to the average value of their initial states instead of its mean square value. This approach addresses continuous-time heterogeneous MASs with time delays, extending existing privacy-preserving methods typically limited to discrete-time systems without time delay.}
\item While the majority of the literature addressing the output regulation problem for heterogeneous MASs assumes that the observers employed be uncompromised to cyber-physical attacks, we remove this strict limitation by developing a fully-distributed bilayer defense framework, which addresses attacks on both CPL and OL. Moreover, the proposed resilient control protocols can effectively handle EU-FDI attacks on both layers. This goes beyond the strict constraint of bounded-first-order-time-derivative attack signals \cite{zuo2023resilient}. Hence, this advancement enriches the capabilities of bipartite output containment control systems in countering more general cyber-physical threats in adversarial environments.

\item A rigorous mathematical proof using Lyapunov stability analysis certifies the UUB consensus and stability of the heterogeneous MASs in the face of EU-FDI attacks, establishing the theoretical soundness of the proposed method. Comparative simulation case studies validate the effectiveness of the proposed bilayer defense strategies.
\end{itemize}

The remainder of this paper is structured as follows: Section II outlines the preliminaries and formulates the problem. Section III presents the design of a fully-distributed attack-resilient defense strategies. Section IV provides validation of the proposed defense strategies through numerical simulations. Finally, Section V conclusions the paper.

\section{Preliminaries and Problem Formulation}

In this section, the preliminaries on graph theory and notations are first given, and then the PABOC problem is formulated.

\subsection{Preliminaries on Graph Theory and Notations}

Consider a group of \( N + M \) agents on a signed communication digraph \( \mathscr{G} \), consisting of \( N \) followers and \( M \) leaders. Leaders are characterized by the absence of incoming edges, thus they operate autonomously. In contrast, followers obtain and process information from their adjacent agents. Denote the follower set and the leader set as $\mathscr{F} = \{v_1, v_2, \ldots, v_N\} \quad \text{and} \quad
\mathscr{L} = \{v_{N+1}, v_{N+2}, \ldots, v_{N+M}\}$ respectively. The interactions among the followers are represented by ${\mathscr{G}_f} = (\mathcal{V},\mathcal{E},\mathcal{A})$ with a nonempty finite set of nodes $\mathcal{V}$, a set of edges $\mathcal{E} \subset \mathcal{V} \times \mathcal{V}$, and \( \mathcal{A} = [a_{ij}] \in \mathbb{R}^{N \times N} \) is the adjacency matrix, where ${a_{ij}}$ is the weight of edge $({v_j},{v_i})$, with \( a_{ij} \neq 0 \) if \( (v_j, v_i) \in \mathcal{E} \); otherwise, \( a_{ij} = 0 \). It is assumed there are no repeated edges and no self-loops, i.e., \( a_{ii} = 0 \), \( \forall i \). A sequence of successive edges in the form $\{ ({v_i},{v_k}),({v_k},{v_l}), \ldots ,({v_m},{v_j})\} $ is a directed path from node $i$ to node $j$. The matrix \( \mathcal G_r = \operatorname{diag}(g_{ir}) \in \mathbb{R}^{N \times N} \), with \( i \in \mathscr{F} \) and \( r \in \mathscr{L} \), represents the diagonal matrix of pinning gains from the \( r \)th leader to each follower. \( g_{ir}\neq0\) if a link from the \( r \)th leader to the \( i \)th follower exists; otherwise, $g_{ir}=0$. It is assumed that the signed digraph \( \mathscr{G} \) is time-invariant, i.e., both \( \mathcal A \) and \( \mathcal G_r \) are constant. 

In this paper, we use the features of global graph topology matrices of two correlated digraphs:
\begin{itemize}
\item[(i)] For the non-negative digraph \( \bar{\mathscr{G}} \), we define the adjacency matrix as \( \bar {\mathcal{A}} = [|{a}_{ij}|] \in \mathbb{R}^{N \times N} \) and the pinning gain matrix as \( \bar{{\mathcal G}}_k = \operatorname{diag}(|g_{ir}|) \in \mathbb{R}^{N \times N} \). The conventional Laplacian matrix is defined as 
$$\bar{\mathcal{L}}= \bar{\mathcal{D}} - \bar{\mathcal{A}} = \operatorname{diag}\left(\sum_{j \in \mathscr{F}} |a_{ij}|\right) - [|a_{ij}|].$$
\end{itemize}
\begin{enumerate}
    \item[(ii)] For the signed digraph \( \mathscr{G} \), consider the adjacency matrix \( \mathcal A = [a_{ij}] \in \mathbb{R}^{N \times N} \) and the matrix of pinning gains \( \mathcal G_r = \operatorname{diag}(g_{ir}) \in \mathbb{R}^{N \times N} \). The signed Laplacian matrix is defined as $$\mathcal{L}^s= \bar{\mathcal{D}} - \mathcal{A} = \operatorname{diag}\left(\sum_{j \in \mathscr{F}} |a_{ij}|\right) - [a_{ij}].$$
\end{enumerate}
Throughout this study, we adopt the following notations:
\begin{itemize}
 %   \item The spectral radius of matrix \( X \) is denoted by \( \rho(X) \).
    \item ${I_N} \in {\mathbb{R}^{N \times N}}$ is the identity matrix. 
    \item ${{\mathbf{1}}_N} \in {\mathbb{R}^N}$ and ${{\mathbf{0}}_N} \in {\mathbb{R}^N}$ are  column vectors with all elements of one and zero, respectively.
    \item The Kronecker product is represented by \( \otimes \).
    \item The operator \( \operatorname{diag}(\cdot) \) is used to form a block diagonal matrix from its argument.
    \item ${\sigma_{\min}}(X)$, ${\sigma _{\max }}(X)$, and $\sigma (X)$ are the minimum singular value, the maximum singular value, and the spectrum of matrix $X$, respectively.
    \item $\left\|  \cdot  \right\|$ is the Euclidean norm of a vector.
\end{itemize}

\subsection{Problem Formulation}
Consider a group of $N$ followers with the following general high-order linear heterogeneous dynamics
\begin{equation}
\label{eq: follower dynamics}
\begin{gathered}
\left\{ \begin{gathered}
  {{\dot x}_i} = {A_i}{x_i} + {B_i}{u^c_i}, \hfill \\
  {y_i} = {C_i}{x_i}, \hfill
\end{gathered}  \right.\quad i \in \mathscr{F}\hfill
\end{gathered}
\end{equation}
where ${x_i} \in \mathbb{R}^{n_i}$ and ${y_i} \in {\mathbb{R}^z}$ are the state and output of the $i$th follower, respectively. ${u_i^c}$$ \in {\mathbb{R}^{m_i}}$ is the compromised input of the $i^{th}$ follower. The local input is under unknown and unbounded actuator attack described by
\begin{equation}
{u_i^c} = {u_i} + {\gamma^a_i},
\label{eq: compromised input}
\end{equation}
where ${u_i} \in \mathbb{R}^{m_i}$ is 
% the intact control input and ${\gamma^a_i}\in \mathbb{R}^{n_i}$ is of class $C^1$ \cite{widder2012advanced} that represents the
EU-FDI attack signal injected to the $i^{th}$ follower\cite{zhu2023secure,li2019event}. The $M$ leaders with the following dynamics can be viewed as command generators that generate the desired trajectories
\begin{equation}
\label{eq: leader dynamics}
\begin{gathered}
\left\{ \begin{gathered}
  {{\dot x }_r} = S{x _r}, \hfill \\
  {y_r} = R{x _r}, \hfill \\ 
\end{gathered}  \right.\quad r \in \mathscr{L}\hfill \\ 
\end{gathered}
\end{equation}
where ${x _r} \in {\mathbb{R}^l}$ and ${y_r} \in {\mathbb{R}^z}$ are the state and output of the $r$th leader, respectively. Noting that $({A_i},{B_i},{C_i})$ and $(S,R)$ may have different system matrices and state dimensions, and hence are heterogeneous.
\begin{remark}
This system modeling is particularly relevant for swarms involving heterogeneous UAVs, including fixed-wing drones, rotary-wing drones (e.g., quadcopters), and hybrid UAVs, each with distinct dynamics and control characteristics tailored for specific operational purposes. The heterogeneity in system matrices and state dimensions (subscripts of \(A_i\), \(B_i\), and \(C_i\)) captures the practical reality of deploying diverse UAV types in collaborative missions, such as search and rescue, environmental monitoring, and surveillance \cite{tang2023swarm}. Consequently, the system matrices  for followers and  for leaders may vary significantly in structure and state dimensions.
In contrast, leaders in formula \eqref{eq: leader dynamics} are modeled with uniform system matrices \(S\) and \(R\) to reflect their advanced and standardized design. This distinction underscores the system's heterogeneity, where diverse followers operate under the guidance of uniform leaders to achieve collaborative objectives efficiently.
\end{remark}

\begin{definition}[Structurally balanced\cite{valcher2014consensus}]
\label{def: structurally balanced}
The signed subgraph ${\mathscr{G}_f}$ is said structurally balanced if it admits a bipartition of the nodes ${\mathcal{V}_1}$, ${\mathcal{V}_2}$, ${\mathcal{V}_1} \cup {\mathcal{V}_2} = \mathcal{V}$, ${\mathcal{V}_1} \cap {\mathcal{V}_2} = 0$, such that ${a_{ij}} \geqslant 0,\forall {v_i},{v_j} \in {\mathcal{V}_q},\left( {q \in \left\{ {1,2} \right\}} \right)$, and ${a_{ij}} \leqslant 0,\forall {v_i} \in {\mathcal{V}_q},{v_j} \in {\mathcal{V}_r},q \ne r,\left( {q,r \in \left\{ {1,2} \right\}} \right)$. It is said structurally unbalanced otherwise.
\end{definition}

\begin{definition}[Convex hull\cite{rockafellar2015convex}]
\label{def: convex hull}
A set $\mathfrak{C} \subseteq {\mathbb{R}^n}$ is convex if $( {1 - \lambda } )x + \lambda y \in \mathfrak{C}$, for any $x,y \in \mathfrak{C}$ and any $\lambda  \in \left[ {0,1} \right]$. Let ${Y_{\mathscr{L}}} = \left\{ {{y_{N + 1}},-{y_{N + 1}},{y_{N + 2}},-{y_{N + 2}},...,{y_{N+M}},-{y_{N + M}}} \right\}$ be the set of the outputs and the negative outputs of the leaders. The convex hull ${\text{Co}}( {Y_{\mathscr{L}}})$ spanned by the outputs and the negative outputs of the leaders is the minimal convex set containing all points in ${Y_{\mathscr{L}}}$. That is, ${\text{Co}}( {{Y_{\mathscr{L}}}} ) = \left\{ {\left. {\sum\limits_{r = N + 1}^{N + M} {({a_r}{y_r}-{b_r}{y_r}} )} \right|{a _r},{b _r} \geqslant 0,\sum\limits_{r = N + 1}^{N + M} {({a_r}+{b_r})}  = 1} \right\}$, where ${\sum_{r = N + 1}^{N + M} {({a_r}{y_r}-{b_r}{y_r}} )}$ is the convex combination of the outputs and the negative outputs of the leaders. 
\end{definition}
\begin{definition}[Distance]
\label{def: distance}
The distance from \( x \in \mathbb{R}^n \) to the set \( \mathcal{C} \in \mathbb{R}^n \) in the sense of Euclidean norm is denoted by $\operatorname{dist}(x, \mathcal{C})$, i.e., $\operatorname{dist}(x, \mathcal{C}) = \inf_{y \in \mathcal{C}} \| x - y \|_2$.
\end{definition}

\begin{definition}[UUB\cite{khalil2002nonlinear}]
\label{def: UUB}
The signal $x(t)\in {\mathbb{R}^n}$ is said to be UUB with the ultimate bound $b$, if there exist positive constants $b$ and $c$, independent of ${t_0} \geqslant 0$, and for every $a \in \left( {0,c} \right)$, there is $T = T\left( {a,b} \right) \geqslant 0$, independent of $t_0$, such that
\begin{equation}
\label{eq12}
\left\| {x\left( {{t_0}} \right)} \right\| \leqslant a\;\; \Rightarrow \;\;\left\| {x\left( t \right)} \right\| \leqslant b,\forall t \geqslant {t_0} + T
\end{equation}
\end{definition}

We have the following assumptions on the communication digraph and the MASs.
\begin{assumption}
\label{ass: leader follower}
Each follower in the
signed digraph ${\mathscr{G}}$, has a directed path from at least one leader.
\end{assumption}

\begin{assumption}
\label{ass: eig of S}
$S$ has non-repeated eigenvalues on the imaginary axis.
\end{assumption}

\begin{assumption}
\label{ass: structurally balanced}
The signed subdigraph ${\mathscr{G}_f}= (\mathcal{V},\mathcal{E},\mathcal{A})$ is structurally balanced.   
\end{assumption}

\begin{assumption}
\label{ass: stabilizable and detectable}
$({A_i},{B_i})$ is stabilizable and $({A_i},{C_i})$ is detectable for each $i \in \mathscr{F}$. 
\end{assumption}   

\begin{assumption}
\label{ass: matrix rank}
\begin{equation}
\label{eq: matrix rank}
\operatorname{rank}\left[ {\begin{array}{*{20}{c}}
  {{A_i} - \lambda {I_{{n_i}}}}&{{B_i}} \\ 
  {{C_i}}&0 
\end{array}} \right] = {n_i} + z,\; \forall \lambda  \in \sigma (S),\;i \in \mathscr{F}.
\end{equation} 
\end{assumption} 

%\begin{definition}
%\label{def: exponentially unbounded signal}
%A signal $\gamma(t)\in {\mathbb{R}^n}$ is said to be exponentially unbounded if $\gamma(t)=+-\exp(\kappa t)$. For the convenience of stability analysis, assume there exists a positive constant $\kappa$, then $\|\gamma(t)\| \leqslant \exp(\kappa t)$.
%\end{definition}

\begin{remark}
\label{rem: remark on the assumptions} Assumption \ref{ass: eig of S} is made to avoid the trivial case when $S$ has eigenvalues with negative real parts. Assumption \ref{ass: stabilizable and detectable}\cite{lewis2013cooperative} and Assumption \ref{ass: matrix rank} \cite{huang2004nonlinear} are standard for output regulation of heterogeneous MASs.
\end{remark}

The following lemmas facilitate the stability analysis of the main result to be presented in the next section.
\begin{lemma}[\cite{valcher2014consensus}]
\label{le: structurally balanced}
Consider the signed subdigraph \( \mathscr{G}_f \). We represent the set of signature matrix set as

\[ \mathcal{Q} = \{ \operatorname{diag}(\sigma_i) \mid \sigma_i \in \{+1, -1\} \}. \]

\( \mathscr{G}_f\) is called structurally balanced if and only if
\begin{enumerate}
    \item The associated undirected graph \( \mathscr{G} (\mathcal A_u) \) is structurally balanced, where \( \mathcal A_u = \frac{\mathcal A + \mathcal A^{\top}}{2} \).
    \item There exists a matrix \( Q = Q^{\top} = Q^{-1} \in \mathcal{Q} \), such that \( \bar {\mathcal A} = [|a_{ij}|] =Q\mathcal AQ \).
\end{enumerate}
\end{lemma}
\begin{lemma}[\cite{zuo2018bipartite}]
\label{le: positive real part eighenvalues and non-negative matrices}
Given Assumption \ref{ass: leader follower} and Assumption \ref{ass: structurally balanced}, denote

\[ \bar \Phi_r = \frac{1}{M} \bar {\mathcal{L}} + \bar{\mathcal{G}}_r, \quad \Phi_r^s = \frac{1}{M} \mathcal{L}^s + \bar{\mathcal{G}}_r. \]

From Lemma \ref{le: structurally balanced}, \( \bar{\mathcal A} = Q \mathcal A Q \), \( \bar {\mathcal D} = Q \bar \Phi_r Q \), \( \bar{\mathcal L} = Q \mathcal{L}^s Q \), and \( \bar {\Phi}_r = Q \Phi_r^s Q \). Thus, \( \bar {\Phi}_r \) and \( \Phi_r^s \) have the same eigenvalues. Therefore, the properties of \( \bar \Phi_r \) and $\sum_{r \in \mathscr{L}}\bar \Phi_r$ in Lemma $7$ in \cite{haghshenas2015containment} also hold for \( \Phi_r^s \) and $\sum_{r \in \mathscr{L}}\Phi_r^s$, that is, \( \Phi_r^s \) and $\sum_{r \in \mathscr{L}}\Phi_r^s$ are positive-definite and nonsingular M-matrices. The following properties hold for both matrices.

\begin{itemize}
    \item[(i)] The eigenvalues of $\Phi_r^s$ and $\sum_{r \in \mathscr{L}} \Phi_r^s$ have positive real parts.
    \item[(ii)] $(\Phi_r^s)^{-1}$ and $\left( \sum_{r \in \mathscr{L}} \Phi_r^s \right)^{-1}$ exist and both are non-negative \cite{haghshenas2015containment}.
\end{itemize}
\end{lemma}

\begin{lemma}[\cite{huang2004nonlinear}]
\label{le: output solution}
Under Assumption 4, the following local output regulator equations have unique solution pairs $\left( {{\Pi _i},{\Gamma _i}} \right)$
\begin{equation}
\label{eq: output solution}
\begin{gathered}
  {A_i}{\Pi _i} + {B_i}{\Gamma _i} = {\Pi _i}S, \hfill \\
  {C_i}{\Pi _i} = R. \hfill \\ 
\end{gathered}
\end{equation}
\end{lemma}

We now introduce the concept of mask function. Consider a continuously differentiable time-varying mask function
\begin{equation}
h : \mathbb{R}_+ \times \mathbb{R}^n \times \mathbb{R}^m \to \mathbb{R}^n
\label{eq: mask function}
\end{equation}
\[
(t, x, \mathfrak{p}) \mapsto h(t, x, \mathfrak{p})
\]
where $\mathfrak{p} \in \mathbb{R}^m$ is a vector of parameters split into $n$ subvectors (not necessarily of the same dimension), one for each node of the network: $\mathfrak{p} = \{\mathfrak{p}_1, \ldots, \mathfrak{p}_n\}$. After applying the mask function, the state $x$ of the system becomes $\breve{x}=h(t, x, \mathfrak{p})$.

\begin{definition}[\cite{altafini2019dynamical}]
\label{def: indiscernible initial condition}
An initial condition $x_0$ is said to be indiscernible from the masked trajectory $\breve x(t)$ if knowledge of the map $h(t, x, \mathfrak{p})$, $t \in [t_0, \infty)$, and the system dynamics of the followers and leaders (see \eqref{eq: follower dynamics} and \eqref{eq: leader dynamics}) is not enough to reconstruct $x_0$. It is said to be discernible otherwise.
\end{definition}

\begin{lemma}[\cite{altafini2019dynamical}]
\label{le: conditions on discernible initial states}
In order to have discernible initial states, the following three conditions must all be satisfied:

(i) The exact functional form of the map $h(\cdot)$ must be known;

(ii) The parameters $\mathfrak{p}$ must be identifiable given the trajectory $h(t, x, \mathfrak{p})$ and the system dynamics \eqref{eq: follower dynamics} and \eqref{eq: leader dynamics};

(iii) The system dynamics of the followers and the leaders (see \eqref{eq: follower dynamics} and \eqref{eq: leader dynamics}) must be observable.

Failure to satisfy (i) and (ii) (or even just (ii)) is enough to guarantee indiscernibility.
\end{lemma}

In order to obfuscate an agent monitoring the communications, the mapping also needs to avoid mapping neighborhoods of a point $x^*$ of \eqref{eq: follower dynamics} and \eqref{eq: leader dynamics} (typically an equilibrium point) into themselves.

\begin{definition}[\cite{altafini2019dynamical}]
\label{def: neiborhood mapping}
A $C^1$ map $h$ is said not to preserve neighborhoods of a point $x^*$ if for all small $\epsilon > 0$, $\|x_0 - x^*\| < \epsilon$ does not imply $\|h(0, x_0, \mathfrak{p}) - x^*\| < \epsilon$.
\end{definition}

\begin{definition}[\cite{altafini2019dynamical}]
\label{def: privacy mask}
The function $h_i(t, x_i, \frak{p}_i)$  is said to be a vanishing privacy mask for agent $i$, if it is local and also satisfies the following conditions
\begin{itemize}
    \item[C1:] $h_i(0, x_i, \frak{p}_i) \neq x_i \quad \forall x_i \in \mathbb{R}^n, \; i = 1, 2, \dots, N;$
    \item[C2:] $h_i(t, x_i, \frak{p}_i)$ guarantees indiscernibility of the initial conditions;
    \item[C3:] $h_i(t, x_i, \frak{p}_i)$ does not preserve neighborhoods of any $x_i \in \mathbb{R}^n$;
    \item[C4:] $h_i(t, x_i, \frak{p}_i)$ strictly increases in $x_i$ for each fixed $t$ and $\frak{p}_i, \; i = 1, 2, \dots, N;$
    \item[C5:] $|h_i(t, x_i, \frak{p}_i) - x_i|$ is decreasing in $t$ for each fixed $x_i$ and $\frak{p}_i$, and $\lim_{t \to \infty} h_i(t, x_i, \frak{p}_i) = x_i, \; i = 1, 2, \dots, N.$
\end{itemize}
\end{definition}

Next, we introduce the PABOC problem for heterogeneous MASs.

\begin{problem}[Privacy-preserving attack-resilient bipartite output containment problem]
\label{pro: paboc}
For the heterogeneous MAS described in \eqref{eq: follower dynamics} and \eqref{eq: leader dynamics} under EU-FDI attacks, the PABOC problem is to design a control input $u_i$ in \eqref{eq: follower dynamics}, and a mask function $h$ in \eqref{eq: mask function}, such that:

(i) the output of each follower converges to a small neighborhood around or within the dynamic convex hull spanned by the outputs and the negative outputs of the leaders. That is, for all initial conditions, $\operatorname{dist}(y_i, \text{Co}({Y_{\mathscr{L}}})), i \in \mathscr{F}$ is UUB.

(ii) the privacy of the data transmitted and/or exchanged is preserved, in the presence of potential eavesdropping. That is, the function $h$ in \eqref{eq: mask function} satisfies the conditions in Definition~\ref{def: privacy mask}.
\end{problem}

\begin{remark}
Swarm systems of heterogeneous UAVs are increasingly studied due to their capability to execute complex tasks through collective behavior. In such systems, a common framework involves leaders and followers interacting on a signed digraph, where leaders generate reference trajectories and followers aim to achieve output containment within the dynamic convex hull spanned by the leaders. The signed digraph structure models both cooperative and antagonistic interactions among agents, capturing practical scenarios where agents may exhibit collaborative behavior or antagonistic tendencies. Additionally, the concept of safe regions is often incorporated to ensure the swarm operates within predefined boundaries, which is critical for avoiding collisions or operating in constrained environments. However, the security of such systems is increasingly challenged by cyberattacks, such as malicious alterations or data spoofing, which can compromise the integrity and reliability of the swarm's operation. Designing resilient control strategies to counteract these attacks and maintain containment under such threats is a pressing research challenge in this domain.    
\end{remark}

To facilitate the stability analysis, we define the following neighborhood bipartite output containment error
\begin{equation}
\begin{gathered}
e^s_{y_i} \equiv \sum_{j \in \mathscr{F}} \left( a_{ij} y_j - |a_{ij}| y_i \right) + \sum_{r \in \mathscr{L}} \left( g_{ir} y_r - |g_{ir}| y_i \right).
\label{eq: def: neighborhood bipartite output containment error}
\end{gathered}
\end{equation}

The next lemma shows that the PABOC problem is solved by ensuring $e^s_{y_i}$ is UUB.

\begin{lemma}
\label{le: the solution of the PABOC problem} 
Under Assumption \ref{ass: leader follower} and Asssumption \ref{ass: structurally balanced}, considering the heterogeneous MAS \eqref{eq: follower dynamics} and \eqref{eq: leader dynamics}, the condition (i) in the PABOC problem is guaranteed if $e^s_{y_i}$ is UUB.
\end{lemma}

\noindent\textbf{\textit{Proof:}}
The neighborhood bipartite output containment error $e^s_{y_i}$ in \eqref{eq: def: neighborhood bipartite output containment error}  can be reformulated as
\begin{equation}
 \begin{gathered}
e^s_{y_i} =\sum_{r \in \mathscr{L}}g_{ir} y_r - \Big(\sum_{j \in \mathscr{F}}|a_{ij}| y_i - \sum_{j \in \mathscr{F}}a_{ij} y_j + \sum_{r \in \mathscr{L}}|g_{ir}| y_i\Big).\hfill
\label{eq: derivation: Bipartite Containment Error}
\end{gathered}   
\end{equation}
Its global form is
\begin{equation}
    \begin{gathered}
    e^s_y = \sum_{r \in \mathscr{L}} \left( \mathcal{G}_r \otimes I_z \right) \left( {\mathbf{1}}_N \otimes y_r \right) - \Big( \left( \mathcal{L}^s \otimes I_z \right)\hfill\\ 
+\sum_{r \in \mathscr{L}} \left( \bar{\mathcal{G}}_r \otimes I_z \right) \Big) y\hfill\\
= \sum_{r \in \mathscr{L}} \left( \mathcal{G}_r \otimes I_z \right) \left( {\mathbf{1}}_N \otimes y_r \right) - \sum_{r \in \mathscr{L}} \bigg( \left( \frac{1}{M} \mathcal{L}^s + \bar{\mathcal{G}}_r \right) \hfill\\\otimes I_z \bigg)
y,\hfill
    \label{eq: global form: Neighborhood Bipartite Containment Error}
    \end{gathered}
\end{equation}
where $e^s_y = {[ {e_{y_1}^{\top},...,e_{y_N}^{\top}} ]^{\top}}$, $y = {[ {y_1^{\top},...,y_N^{\top}} ]^{\top}}$. For convenience, denote $\bar y_r = \mathbf{1}_N \otimes y_r$. Note that $\left( \bar{\mathcal{L}} \otimes I_z \right) \left( \mathbf{1}_N \otimes y_r \right) = 0,\:\forall r \in \mathscr{L}$. Further manipulation of equation \eqref{eq: global form: Neighborhood Bipartite Containment Error} yields
\begin{equation}
    \begin{gathered}
    e^s_{y} = \sum_{r \in \mathscr{L}} \left(\left( \frac{1}{M} \bar{\mathcal{L}} + \mathcal{G}_r \right) \otimes I_z \right)\bar y_r - \sum_{r \in \mathscr{L}} \left( \Phi_r^s \otimes I_z \right) y \hfill \\
    = -\sum_{\nu \in \mathscr{L}} (\Phi_{\nu}^s \otimes I_z) \Bigg( y - \hfill\\
    \left( \sum_{k \in \mathscr{L}} (\Phi_k^s \otimes I_z) \right)^{-1} \left( \sum_{r \in \mathscr{L}} \left(\frac{1}{M} \bar{\mathcal{L}} + \mathcal{G}_r \right) \otimes I_z \right)\bar{y}_r \Bigg) \hfill\\
    = -\sum_{\nu \in \mathscr{L}} (\Phi_{\nu}^s \otimes I_z) \Bigg( y - \hfill\\
    \sum_{r \in \mathscr{L}} \left( \left( \sum_{k \in \mathscr{L}} \Phi_k^s \right)^{-1} \otimes I_z \right) \left(\left(\frac{1}{M} \bar{\mathcal{L}} + \mathcal{G}_r \right) \otimes I_z\right) \bar{y}_r\Bigg)\\
    = -\sum_{\nu \in \mathscr{L}} (\Phi_{\nu}^s \otimes I_z) \Bigg(y -\hfill\\
    \sum_{r \in \mathscr{L}} \Bigg( \left( \sum_{k \in \mathscr{L}} \Phi_k^s  \right)^{-1} \left( \frac{1}{M} \bar{\mathcal{L}} + \mathcal{G}_r \right) \mathbf{1}_N \Bigg) \otimes y_r\Bigg).\hfill
    \end{gathered}
\end{equation}

Let
% \begin{equation}
% \left\{
% \begin{aligned}
% \mathcal{M}_r &= \frac{1}{M} \bar{\mathcal{L}} + \frac{1}{2M} (\bar{\mathcal{A}} - \mathcal A) + \frac{1}{2} (\bar{\mathcal{G}}_r + \mathcal{G}_r), & r \in \mathscr{L}\\
% \mathcal{N}_r &= \frac{1}{2M} (\bar{\mathcal{A}} - \mathcal A) + \frac{1}{2} (\bar{\mathcal{G}}_r - \mathcal{G}_r), & r \in \mathscr{L}
% \end{aligned}
% \right.
% \end{equation}
\begin{equation}
\begin{gathered}
\left\{\begin{gathered}
\mathcal{M}_r = \frac{1}{M} \bar{\mathcal{L}} + \frac{1}{2M} (\bar{\mathcal{A}} - \mathcal A) + \frac{1}{2} (\bar{\mathcal{G}}_r + \mathcal{G}_r),\\\hfill 
\mathcal{N}_r = \frac{1}{2M} (\bar{\mathcal{A}} - \mathcal A) + \frac{1}{2} (\bar{\mathcal{G}}_r - \mathcal{G}_r),\hfill
\end{gathered}\right. \quad r \in \mathscr{L}\hfill
\end{gathered}
\end{equation}

We obtain
\begin{equation}
\begin{gathered}
e^s_y = -\sum_{\nu \in \mathscr{L}} \left( \Phi^s_{\nu}  \otimes I_z \right) 
\Bigg( y - \sum_{r \in \mathscr{L}} \Bigg(\left( \sum_{k \in \mathscr{L}} \Phi^s_k \right)^{-1}\hfill\\
\times\left( \mathcal{M}_r - \mathcal{N}_r \right) {\mathbf{1}}_N \Bigg)\otimes y_r \Bigg).\hfill
\label{eq: e^s_y further manipulated}
\end{gathered}
\end{equation}

Next, we prove that $\sum_{r \in \mathscr{L}}\Big(\left(\sum_{k \in \mathscr{L}} \Phi^s_k\right)^{-1} ( \mathcal{M}_r + \mathcal{N}_r )\mathbf{1}_N\Big)  ={\mathbf{1}}_N$, meaning that, each element of the column vector, formed by summing $\left(\sum_{k \in \mathscr{L}} \Phi^s_{k}\right)^{-1} ( \mathcal{M}_r + \mathcal{N}_r )\mathbf{1}_N$, is $1$. The proof follows.

\begin{equation}
\begin{gathered}
\sum_{r \in \mathscr{L}} \left( \left( \sum_{k \in \mathscr{L}} \Phi^s_k \right)^{-1} (\mathcal{M}_r + \mathcal{N}_r) {\mathbf{1}}_N \right)\\
= \left( \sum_{k \in \mathscr{L}} \Phi^s_k \right)^{-1} \left( \sum_{r \in \mathscr{L}} \left( \frac{1}{M} \mathcal{L}^s + \bar{\mathcal{G}}_r \right) {\mathbf{1}}_N \right) \\
= \left( \sum_{k \in \mathscr{L}} \Phi^s_k \right)^{-1} \left( \sum_{r \in \mathscr{L}} \Phi^s_r {\mathbf{1}}_N \right) = {\mathbf{1}}_N.
\end{gathered}
\end{equation}
Subsequently, our analysis confirms that every element within the vectors $\left(\sum_{k \in \mathscr{L}} \Phi^s_k\right)^{-1} \mathcal{M}_r {\mathbf{1}}_N$ and $\left(\sum_{k \in \mathscr{L}} \Phi^s_k\right)^{-1} \mathcal{N}_r {\mathbf{1}}_N$, $r \in \mathscr{L}$ is non-negative. We know that $\bar{\mathcal{L}} {\mathbf{1}}_N = \mathbf{0}_N$. Given that the matrices $(\bar{\mathcal{A}} - \mathcal A)$ and $(\bar{\mathcal{G}}_r + \mathcal{G}_r)$ are non-negative, and referring to Lemma \ref{le: positive real part eighenvalues and non-negative matrices}, we find that the matrix $\left(\sum_{k \in \mathscr{L}} \Phi^s_k\right)^{-1}$ exists and is non-negative. Therefore, we obtain that the vector $\left(\sum_{k \in \mathscr{L}} \Phi^s_k\right)^{-1} \mathcal{M}_r {\mathbf{1}}_N$, $r \in \mathscr{L}$ is non-negative. Similarly, we obtain that $\left(\sum_{k \in \mathscr{L}} \Phi^s_k\right)^{-1} \mathcal{N}_r {\mathbf{1}}_N$, $r \in \mathscr{L}$, is non-negative. Subsequently, the term $\left(\sum_{r \in \mathscr{L}} \left( \sum_{k \in \mathscr{L}} \Phi^s_k \right)^{-1} (\mathcal{M}_r - \mathcal{N}_r) {\mathbf{1}}_N \otimes y_r \right)$ described in \eqref{eq: e^s_y further manipulated} represents a column vector of the convex combinations of the outputs and negative outputs of the leaders. From Lemma \ref{le: positive real part eighenvalues and non-negative matrices}, $\sum_{r \in \mathscr{L}} \left( \Phi_r \otimes I_z \right)$ is a nonsingular matrix. Hence, $e^s_{y_i}$ is UUB implies that the following is UUB.
\begin{equation}
    \begin{gathered}
    y - \sum_{r \in \mathscr{L}}\left( \left( \sum_{k \in \mathscr{L}} \Phi^s_k \right)^{-1} (\mathcal{M}_r - \mathcal{N}_r){\mathbf{1}}_N \right)
    \otimes y_r.\hfill
    \end{gathered}
    \label{eq: UUB equivalently}
\end{equation}
According to Definition \ref{def: distance}, \eqref{eq: UUB equivalently} is UUB is equivalent to $\operatorname{dist}(y_i, \text{Co}({Y_{\mathscr{L}}})), i \in \mathscr{F}$ is UUB. Hence, the proof is completed.
$\hfill \blacksquare$

\section{Fully-distributed Privacy-preserving Attack-Resilient Bilayer Defense Strategy Design}

\begin{figure}
\centering
\includegraphics[width=3.5in]{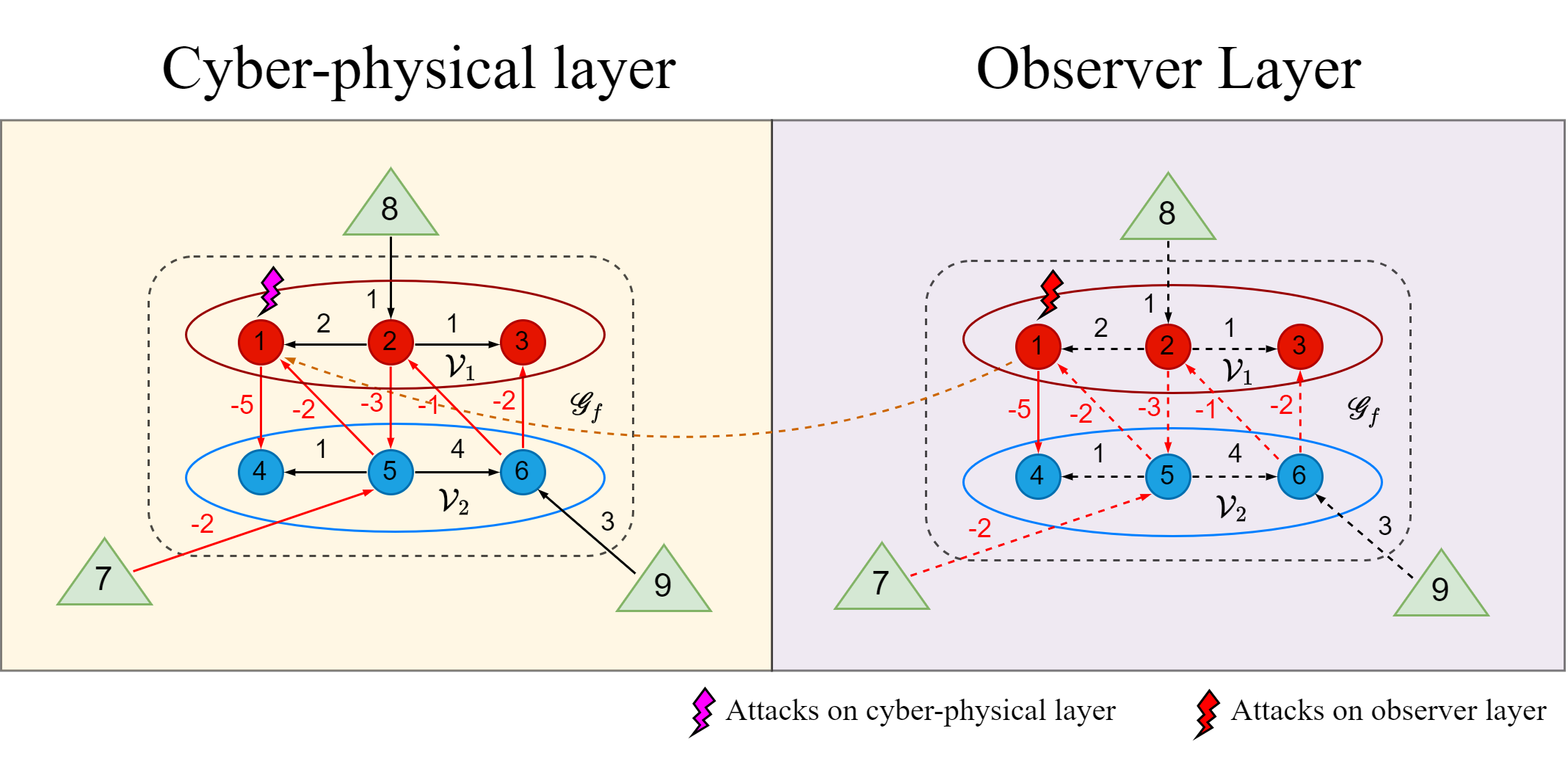}
\captionsetup{justification=centering}
\caption{Cyber-physical layer and observer layer.}
\label{fig: cyber-physical layer and observer layer}
\end{figure}
\begin{figure*}[t]
\centering
\includegraphics[width=1\textwidth]{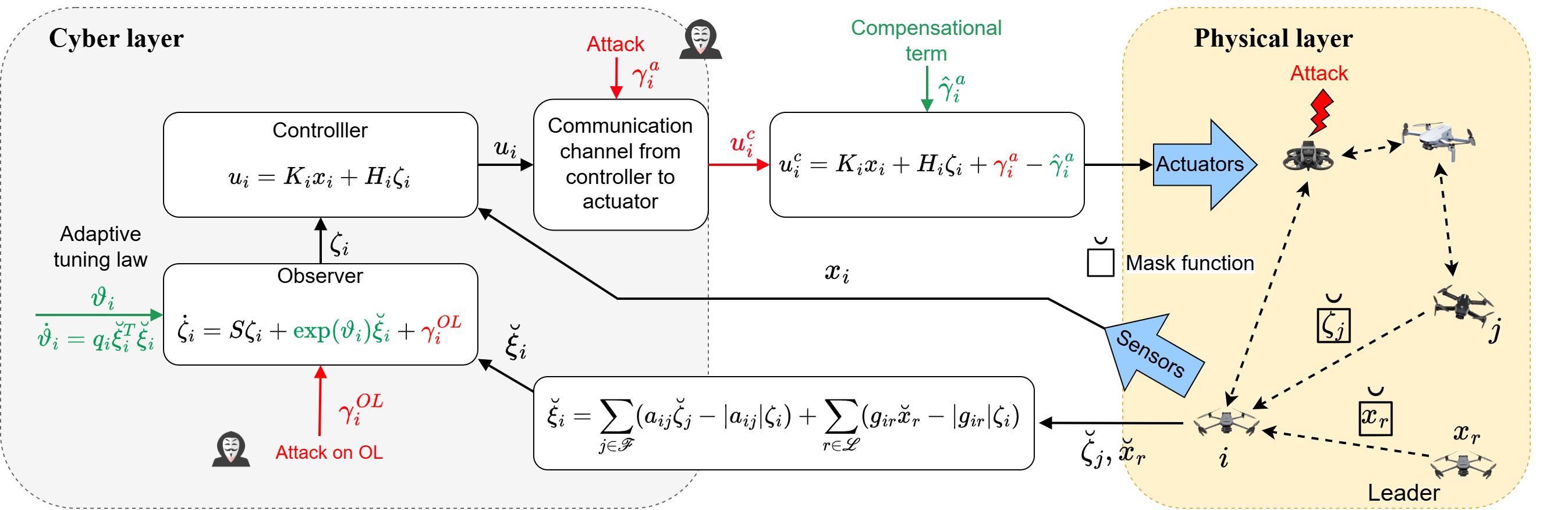}
\captionsetup{justification=centering}
\caption{The overall closed-loop cyber-physical dynamical system.}
\label{fig: cyber-physical system}
\end{figure*}
In this section, the privacy-preservation for heterogeneous
MASs via mask function design is proposed first. A mask function is introduced to hide the states of the leaders and the states transmission among followers on the digital OL, such that the privacy of the initial states is preserved. Next, we develop fully-distributed attack-resilient control strategies to solve the PABOC problem for heterogeneous MASs by using a bilayer defense architecture, as illustrated in Fig.~\ref{fig: cyber-physical layer and observer layer}, where the communication network comprises six followers represented by circles and three leaders represented by triangles. We first construct dynamic compensators communicating on the OL to estimate the convex combinations of the sates and negative states of the leaders. While prevailing literature generally assumes that there is no cyber-attack on the digital OL, we relax such strict limitation by considering the potential cyber-attacks on the digital OL. The information flow among agents are represented by arrows, with the corresponding edge weight values annotated adjacent to them. Positive edge weight values indicate cooperative relationships and negative edge weight values indicate antagonistic relationships. We consider a more practical and challenging scenario where the OL is also subjected to cyber-attacks, necessitating the design of an attack-resilient dynamic compensators.

For convenience, we first define the following neighborhood bipartite state containment error on the OL
\begin{gather}
\xi_i=
\sum_{j \in \mathscr{F}} (a_{ij}\zeta_{j}- |a_{ij}|\zeta_{i})+\sum_{r \in \mathscr{L}}(g_{ir}x_{r} - |g_{ir}|\zeta_{i}),
\label{eq: xii}
\end{gather}
where $\zeta_{i}$ is the local observer state on the OL. As seen, the leaders' states and the observes' states are exchanged on the digital OL. Motivated by \cite{altafini2019dynamical}, to preserve the privacy of the information, the following mask functions are designed.

The function 
\begin{equation}
h(t,x_r, \frak{p}_i)) = \left(1 + \phi^l_ie^{-\sigma^l_it}\right)\left( x_r + \wp^l_i e^{-\delta^l_i t} \right)
\label{eq: the mask function for leader}
\end{equation}
is a mask of privacy in the state $x_r(t)$, where $\phi^l_i>0, \sigma^l_i>0, \delta^l_i>0, \wp^l_i \neq 0$.

The function 
\begin{equation}
h(t,\zeta_j, \frak{p}_j)) = \left(1 + \phi^f_j e^{-\sigma^f_it}\right)\left( \zeta_j + \wp^f_j e^{-\vartheta_j(t)} \right)
\label{eq: our mask function for followers}
\end{equation}
is a mask of privacy in the state $\zeta_j(t)$, where $\phi^f_j>0, \sigma^f_j>0, \delta^f_j>0, \wp^f_j \neq 0$. $\vartheta_j$ is to be designed.

After employing the mask function \eqref{eq: the mask function for leader} and \eqref{eq: our mask function for followers}, the data $x_r$ and $\zeta_j$ transmitted from the leader and the neighboring followers, respectively, in \eqref{eq: xii} becomes

\begin{gather}
\breve\xi_i=
\sum_{j \in \mathscr{F}} (a_{ij}\breve{\zeta}_j- |a_{ij}|\zeta_{i})+\sum_{r \in \mathscr{L}}(g_{ir}\breve x_{r} - |g_{ir}|\zeta_{i}).
\label{eq: masked xii}
\end{gather}
Then, we develop the following fully-distributed attack-resilient dynamical observer against EU-FDI attacks on the OL

\begin{align}
& \dot{\zeta}_i = S\zeta_i + \exp{(\vartheta_i)}{\breve\xi_i} + \gamma^{OL}_i,
\label{eq: zetai dot}\\
&\dot{\vartheta}_i = q_i\breve\xi_i^{\top}{\breve\xi_i},
\label{eq: theta i dot}
\end{align}
where ${\vartheta_i}$ is adaptively tuned by \eqref{eq: theta i dot} with ${\vartheta _i}\left( 0 \right) = 0$, $q_i>0$ is the coupling gain in the adaptive tuning law, and $\gamma^{OL}_i$ denotes the EU-FDI attack signal targeting observer $i$ on the digital OL\cite{yang2023distributed}.

\begin{definition}
\label{def: exponentially unbounded signal}
A signal $\gamma(t)\in {\mathbb{R}^n}$ is said to be exponentially unbounded if $\gamma(t) = [ {k_1\exp(\kappa_1 t),...,k_n\exp(\kappa_n t)} ]^{\top}$, where $\kappa_1$,..., $\kappa_n$ are positive constants and $k_n$ are constant coefficients, which could be unknown.
\end{definition}

\begin{assumption}
\label{ass: attacks}
$\gamma^a_i(t)$ and $\gamma^{OL}_i(t)$ are exponentially unbounded signals.
\end{assumption}

\begin{remark}
\label{rem: compromised observer}
Observer design is generally employed to estimate certain convex combinations of the leaders' states for heterogeneous MASs. However, most of the literature assumes that the digital OL remain intact against cyber-attacks, which is not practical. In contrast, we consider more practical and challenging scenarios in which the observers could also be attacked. Attackers, such as hackers, can inject false data into the system, exploiting vulnerabilities in communication protocols. For instance, man-in-the-middle attacks leverage tampering with the address resolution protocol \cite{zhao2020arp} to intercept, modify, or inject false data during communication \cite{conti2016survey}. To ensure observation validity, we propose an intelligent observer with an adaptive tuning law designed to counteract the effects of false data injections as shown in Fig.~\ref{fig: cyber-physical system}. This design ensures that the estimation error is UUB, maintaining the efficacy of the digital OL even under EU-FDI attacks.
\end{remark}

\begin{remark}
\label{rem: remark on the attack signals}
As described in Assumption \ref{ass: attacks} and shown in the stability analysis in the Appendix, the defense capabilities of the designed attack-resilient controller is significantly expanded, which address a wide range of FDI attack signals, including those that grow exponentially over time. In reality, adversaries can inject any time-varying signal into systems via software, CPU, DSP, or similar platforms. Note that Assumption \ref{ass: attacks} represent the worst-case scenarios that the controller can manage. That is, the proposed controller is capable of handling a broad spectrum of FDI attack signals, compared with \cite{cheng2023adaptive,zuo2023resilient}.
\end{remark}

%Exponentially growing signals are a specific type of time-varying signal. These signals contribute to the versatility of attack signal types that the controller can address.

\begin{remark}
\label{rem: justification of privacy preserving}
The observer design presented in \eqref{eq: masked xii}-\eqref{eq: theta i dot} incorporates a privacy-preserving mechanism by applying a mask function to the data transmitted among agents. Specifically, the observer states, transmitted among followers, and the leaders' states are masked to ensure the confidentiality of critical information during communication. The observer's primary role is to estimate certain convex combinations of the leaders' states, which subsequently determine the trajectories of the followers. Preserving these trajectories is essential in applications such as UAV swarms operating in adversarial environments, where the followers' movements often reflect sensitive mission dynamics and coordination strategies. Unauthorized access to these trajectories could jeopardize operational security and mission success. Additionally, as shown in \eqref{eq: our mask function for followers}, the observer mask function employs a time-varying adaptive parameter $\vartheta_j(t)$ to enhance privacy, making it significantly more challenging to infer the observers' states or the followers' trajectories from intercepted data.
\end{remark}

\begin{remark}
\label{rem: remark on fully-distributed}
In \cite{lewis2013cooperative}, the knowledge of the global graph topology is required to design the coupling gain in the dynamical observer design. However, as seen from Eq. \eqref{eq: theta i dot}, no knowledge of the global graph topology is required in the design of the adaptive coupling gain $\vartheta_i$. Hence, the controller is fully-distributed.
\end{remark}

Define the following state tracking error
\begin{equation}
\label{eq: epsiloni}
{\varepsilon _i} = {x_i} - {\Pi _i}{\zeta _i}.
\end{equation}
Building on the dynamic resilient observer design, we finally introduce the following fully-distributed attack-resilient controller design.
\begin{align}
& {u_i} = {K_i}{x_i} + {H_i}{\zeta _i} - \hat \gamma^a_i,
\label{eq: controller ui}\\
& {\hat \gamma^a_i} = \frac{{B_i^{\top}{P_i}{\varepsilon _i}}}{{\left\| {{\varepsilon _i}^{\top}{P_i}{B_i}} \right\| + \exp \left( - c_it^2 \right)}}\exp(\hat \rho _i),
\label{eq: hat gamma bi}\\
& {{\dot {\hat \rho} }_i} ={\alpha_i\left\| {{\varepsilon _i}^{\top}{P_i}{B_i}} \right\|} ,
\label{eq: dot hat rho i}
\end{align}
where $\hat \gamma^a_i$ is a compensational signal designed per \eqref{eq: hat gamma bi} to mitigate the adverse effect caused by the actuator attack signal $\gamma^a_i$, $\hat \rho_i$ is a gain adaptively tuned by \eqref{eq: dot hat rho i}, $\alpha_i$ and $c_i$ are positive constants. The overall closed-loop cyber-physical dynamical system is illustrated in Fig.~\ref{fig: cyber-physical system}.

Employ certain positive-definite symmetric matrices $U_i$ and $Q_i$, under Assumption \ref{ass: stabilizable and detectable}, the solution $P_i$ to the following algebraic Riccati equation can be found.
\begin{equation}
A_i^{\top} P_i+P_i A_i+Q_i-P_iB_i{U_i^{-1}}{B_i^{\top}}{P_i}=0.
\label{eq: solve for P}
\end{equation}
The controller gain matrices $K_i$ and $H_i$ in \eqref{eq: controller ui} are designed as 
\begin{align}
& {K_i} =  - U_i^{ - 1}B_i^{\top}{P_i},
\label{eq: Ki}\\
& H_i=\Gamma_i-K_i {\Pi_i},
\label{eq: Hi}
\end{align}

% {\color{red}\begin{remark}As illustrated in Figure 2, the proposed framework integrates practical implementation techniques for compensators and controllers to enhance UAV resilience against bounded FDI attacks. Compensators are implemented as algorithmic modules within the control system, leveraging low-latency hardware such as digital signal processors or field-programmable gate arrays and embedding adaptive tuning mechanisms, such as Kalman filters for real-time adjustments. Controllers, including PID or model predictive controllers, are discretized and optimized for performance through simulations or data-driven approaches, ensuring robustness in sampled-data systems. Under bounded FDI attacks, detection schemes employ state observers or statistical models to isolate corrupted signals, enabling compensators to dynamically re-weight inputs and redistribute control efforts. Meanwhile, controllers adapt control signals to maintain stability and mitigate performance degradation. Figure 2 visually demonstrates the architecture of the closed-loop system, including the compensator and observer's interactions to counteract adversarial effects, bridging theoretical development and practical deployment in lightweight, scalable UAV systems.\end{remark}}

Next, we present the main result for solving the PABOC problem for heterogeneous MASs.

\begin{theorem}
\label{thm: solve pro: paboc}
Given Assumptions \ref{ass: leader follower} to \ref{ass: attacks}, considering the heterogeneous MAS composed of \eqref{eq: follower dynamics} and \eqref{eq: leader dynamics} in the presence of EU-FDI attacks on both CPL and OL, Problem \ref{pro: paboc} is solved by designing the fully-distributed controller consisting of \eqref{eq: xii}-\eqref{eq: Hi} and the mask functions $h$ as designed in \eqref{eq: the mask function for leader} and \eqref{eq: our mask function for followers}.
\end{theorem}

\textit{\textbf{Proof:} See proof of Theorem 1 in the appendix.}

\section{Numerical Simulations}
\begin{figure}[!h]
\centering
\includegraphics[width=2in]{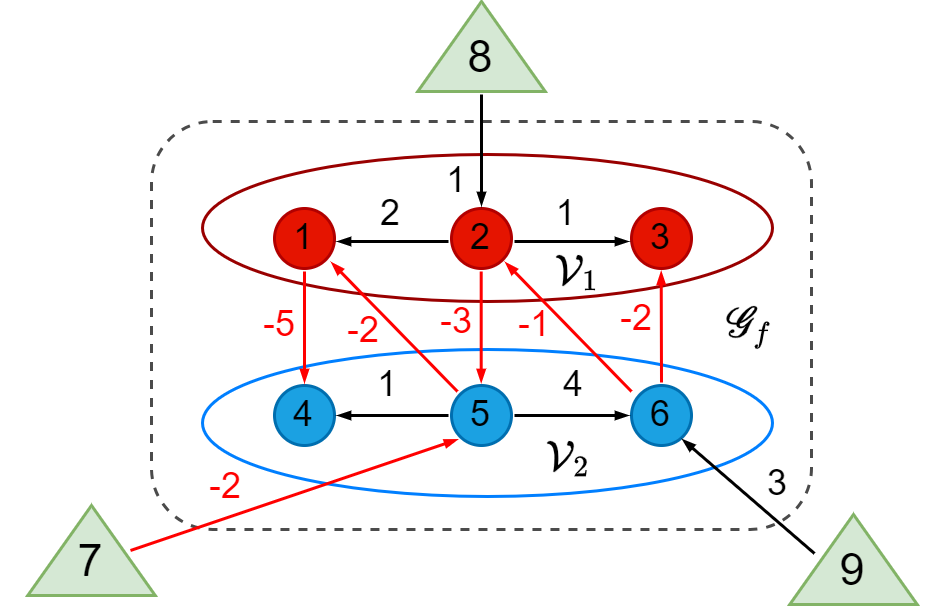}
\captionsetup{justification=centering}
\caption{Communication topology.}
\label{fig: Communication topology}
\end{figure}
%The second study demonstrates the application of our control protocols in quadrotor systems, further confirming the resilience of our protocols when the system is subjected to EU-FDI attack signals. Together, these studies validate the adaptability and resilience of our proposed control protocols in diverse and challenging adversarial environments.
In this section, we validate our proposed cyber-physical defense strategies within a general heterogeneous MAS, specifically verifying the effectiveness and resilience of the control protocols against EU-FDI attack signals in the presence of eavesdroppers. The communication topology of the heterogeneous MAS is delineated in Fig.~\ref{fig: Communication topology}. The system has six circle followers and three triangle leaders. The dynamics of the followers and leaders are given by:

\begin{equation}
\label{eq: leader follower dynamics}
\begin{gathered}
\left\{ \begin{gathered}
  \dot{x}_{1,2} = \left[ \begin{array}{*{2}{c}}
  -2 & 1 \\ 
  0 & -3 
\end{array} \right] x_{1,2} + \left[ \begin{array}{*{2}{c}}
  1 & 0 \\ 
  0 & 1 
\end{array} \right] u_{1,2} \hfill \\
  y_{1,2} = \left[ \begin{array}{*{2}{c}}
  0.5 & 1 \\ 
  1 & 0.5 
\end{array} \right] x_{1,2} \hfill \\ 
\end{gathered}  \right.\hfill \\
\left\{ \begin{gathered}
  \dot{x}_{3,4} = \left[ \begin{array}{*{2}{c}}
  -1 & 0 \\ 
  0 & -2 
\end{array} \right] x_{3,4} + \left[ \begin{array}{*{2}{c}}
  0.5 & 1 \\ 
  1 & 0.5 
\end{array} \right] u_{3,4} \hfill \\
  y_{3,4} = \left[ \begin{array}{*{2}{c}}
  1 & 0.5 \\ 
  0.5 & 1 
\end{array} \right] x_{3,4} \hfill \\ 
\end{gathered}  \right.\hfill \\
\left\{ \begin{gathered}
  \dot{x}_{5,6} = \left[ \begin{array}{*{3}{c}}
  -1 & 0 & 0 \\ 
  0 & -2 & 0 \\ 
  0 & 0 & -3  
\end{array} \right] x_{5,6} + \left[ \begin{array}{*{2}{c}}
  1 & 0 \\ 
  0 & 1 \\ 
  1 & 0 
\end{array} \right] u_{5,6} \hfill \\
  y_{5,6} = \left[ \begin{array}{*{3}{c}}
  1 & 0 & -1 \\ 
  0 & 1 & 1 
\end{array} \right] x_{5,6} \hfill \\ 
\end{gathered}  \right.\hfill \\
\left\{ \begin{gathered}
  \dot{x}_{7,8,9} = \left[ \begin{array}{*{2}{c}}
  0 & -2 \\ 
  1 & 0 
\end{array} \right] x_{7,8,9}, \\
  y_{7,8,9} = \left[ \begin{array}{*{2}{c}}
  1 & 0 \\ 
  0 & 1 
\end{array} \right] x_{7,8,9} \hfill \\
\end{gathered} \right.\hfill \\
\end{gathered}
\nonumber \\
\end{equation}

We choose the following EU-FDI attack signals injected on CPL and OL:

\begin{equation*}
\begin{aligned}
    \gamma^{a}_{1} &= \begin{bmatrix} 2.3 e^{0.12t} \\ -1.7 e^{0.27t} \end{bmatrix}, & \gamma^{OL}_{1} &= \begin{bmatrix} -3.2 e^{0.23t} \\ 2.1 e^{0.45t} \end{bmatrix}, \\
    \gamma^{a}_{2} &= \begin{bmatrix}  3.9 e^{0.08t} \\ -2.5 e^{0.35t} \end{bmatrix}, & \gamma^{OL}_{2} &= \begin{bmatrix} 1.7 e^{0.32t} \\ -3.8 e^{0.16t} \end{bmatrix},\\
    \gamma^{a}_{3} &= \begin{bmatrix} -4.2 e^{0.15t} \\ 1.8 e^{0.24t} \end{bmatrix}, & \gamma^{OL}_{3} &= \begin{bmatrix} -4.5 e^{0.41t} \\ 1.2 e^{0.29t} \end{bmatrix},\\
    \gamma^{a}_{4} &= \begin{bmatrix}  0.9 e^{0.18t} \\ -3.6 e^{0.11t} \end{bmatrix}, & \gamma^{OL}_{4} &= \begin{bmatrix} 2.3 e^{0.37t} \\ -0.9 e^{0.12t} \end{bmatrix}, \\
    \gamma^{a}_{5} &= \begin{bmatrix} -1.5 e^{0.23t} \\ 2.7 e^{0.14t} \end{bmatrix}, & \gamma^{OL}_{5} &= \begin{bmatrix} -0.8 e^{0.21t} \\ 3.4 e^{0.08t} \end{bmatrix}, \\
    \gamma^{a}_{6} &= \begin{bmatrix} 3.4 e^{0.05t} \\ -0.8 e^{0.29t}  \end{bmatrix}, & \gamma^{OL}_{6} &= \begin{bmatrix} 3.9 e^{0.15t} \\ -2.7 e^{0.05t} \end{bmatrix}.
\end{aligned}
\end{equation*}

These exponentially growing attack signals are designed to test the system's resilience and adaptability in dynamic adversarial scenarios. The following pairs $( \Pi _i, \Gamma _i)$ are obtained for each follower by solving \eqref{eq: output solution}
\begin{equation}
\begin{gathered}
  {\Pi _{1,2}} = \left[ {\begin{array}{*{20}{c}}
  {-0.67}&{1.33} \\ 
  {1.33}&{-0.67} 
\end{array}} \right],{\Gamma _{1,2}} = \left[ {\begin{array}{*{20}{c}}
  {-1.33}&{4.67} \\ 
  {3.33}&{-4.67} 
\end{array}} \right], \hfill \\
  {\Pi _{3,4}} = \left[ {\begin{array}{*{20}{c}}
  {1.33}&{-0.67} \\ 
  {-0.67}&{1.33} 
\end{array}} \right],{\Gamma _{3,4}} = \left[ {\begin{array}{*{20}{c}}
  {-0.44}&{7.56} \\ 
  {0.89}&{-7.11} 
\end{array}} \right], \hfill \\
  {\Pi _{5,6}} = \left[ {\begin{array}{*{20}{c}}
  {1.50}&{-1.00} \\ 
  {-0.50}&{2.00} \\ 
  {0.50}&{-1.00} 
\end{array}} \right],{\Gamma _{5,6}} = \left[ {\begin{array}{*{20}{c}}
  {0.50}&{-4.00} \\ 
  {1.00}&{5.00} 
\end{array}} \right]. \hfill \\ 
\end{gathered}
\nonumber \\
\end{equation}

Select $U_{1,2,...,6}=I_2$, $Q_{1,2,3,4}=3I_2$, and $Q_{5,6}=3 I_3$. The controller gain matrices $K_i$ and $H_i$ are found by solving \eqref{eq: Ki} to \eqref{eq: solve for P} are
\begin{equation}
\begin{gathered}
  {K_{1,2}} = \left[ {\begin{array}{*{20}{c}}
  { -0.64}&{ -0.10} \\ 
  {-0.10}&{ -0.49} 
\end{array}} \right],{H_{1,2}} = \left[ {\begin{array}{*{20}{c}}
  {-1.62}&{5.46} \\ 
  {3.92}&{-4.86} 
\end{array}} \right], \hfill \\
  {K_{3,4}} = \left[ {\begin{array}{*{20}{c}}
  {-0.37}&{-0.59} \\ 
  {-0.93}&{-0.19} 
\end{array}} \right],{H_{3,4}} = \left[ {\begin{array}{*{20}{c}}
  {-0.35}&{8.09} \\ 
  {2.00}&{ -7.47} 
\end{array}} \right], \hfill \\
  {K_{5,6}} = \left[ {\begin{array}{*{20}{c}}
  {-0.95}&{0}&{-0.38} \\ 
  {0}&{-0.65}&{0} 
\end{array}} \right], \hfill \\
{H_{5,6}} = \left[ {\begin{array}{*{20}{c}}
  {2.12}&{-5.34} \\ 
  {0.68}&{6.29} 
\end{array}} \right]. \hfill \\ 
\end{gathered}
\nonumber \\
\end{equation}

For comparison, we run the simulation using the standard bipartite output containment control protocols as follows.

\begin{equation}
\begin{gathered}
\left\{\begin{gathered}
{{\dot \zeta }_i} = S{\zeta _i} + \vartheta_i{\xi_i},
\hfill\\
{{\dot \vartheta }_i} = q_i\xi_i^{\top}{\xi_i},
\hfill\\
{u_i} = {K_i}{x_i} + {H_i}{\zeta _i}.\hfill
\end{gathered}\right.
\end{gathered}
\label{eq: standard ctrl prcls}
\end{equation}
Next, we evaluate the system's resilience against EU-FDI attacks on CPL and OL using the standard bipartite output containment control protocols and the proposed cyber-physical defense strategies. The outputs and the negative outputs of the leaders and the outputs of the followers are captured as snapshots at three time instants in both comparative simulation case studies, where the outputs of leaders are denoted by green triangles, and the negative outputs of leaders are denoted by purple triangles. The EU-FDI attacks on CPL and OL are initiated simultaneously at $8$ s.
\begin{figure}[!h]
\centering
\includegraphics[width=9cm]{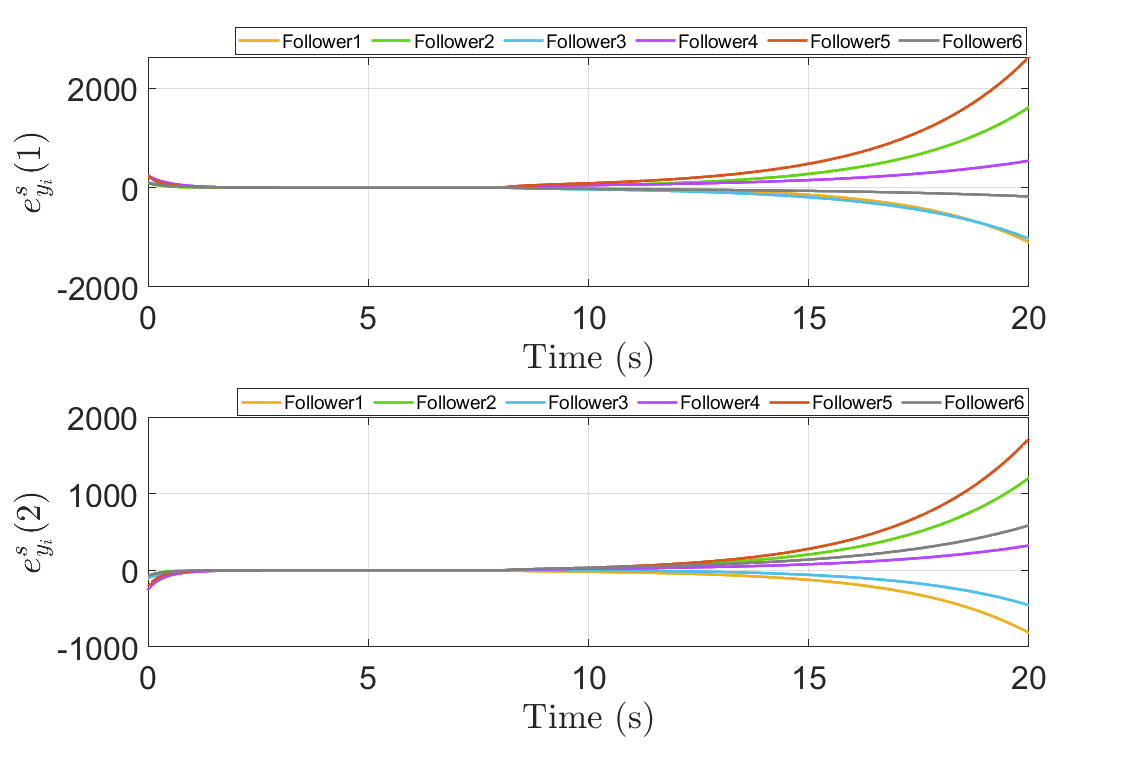}
\captionsetup{justification=centering}
\caption{Bipartite output containment errors $e^s_{y_i}$ using the standard control protocols: $e^s_{y_i}(1)$ is the $x$ coordinate of $e^s_{y_i}$, $e^s_{y_i}(2)$ is the $y$ coordinate of $e^s_{y_i}$.}
\label{fig: Errors without resilient controller}
\end{figure}
\begin{figure}[!h]
\centering
\includegraphics[width=9cm]{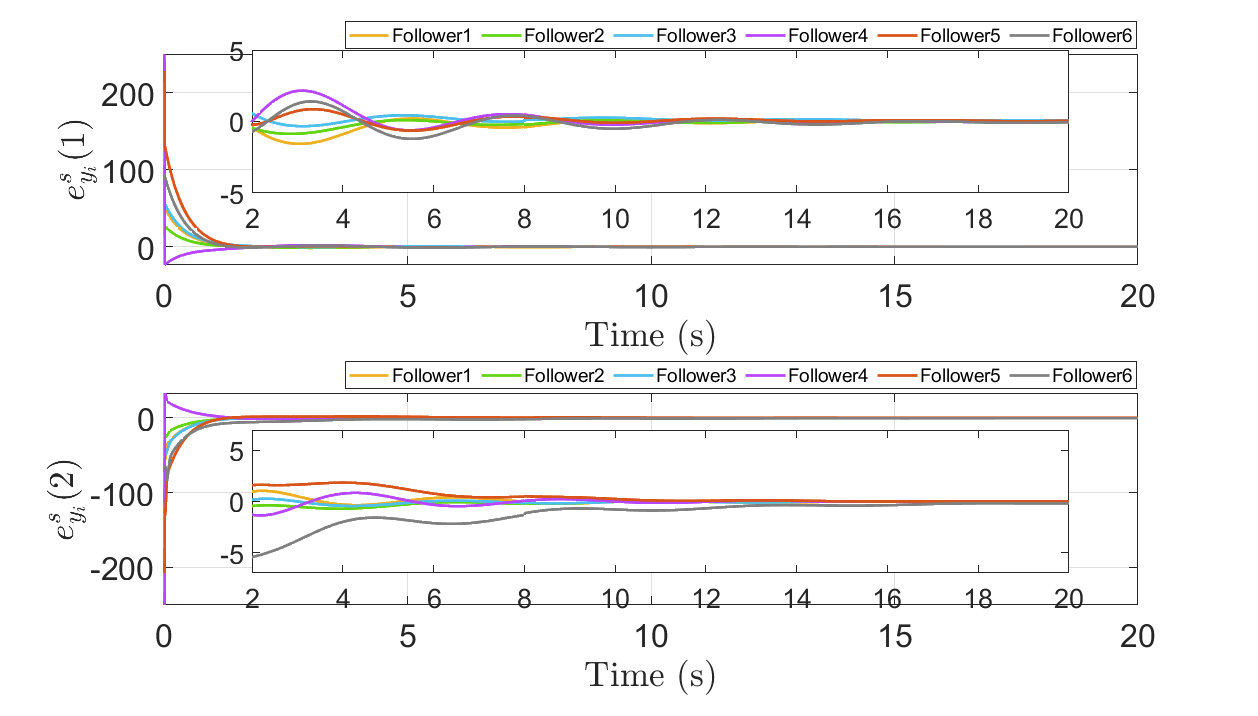}
\captionsetup{justification=centering}
\caption{Bipartite output containment errors $e^s_{y_i}$ using the proposed resilient control protocols: $e^s_{y_i}(1)$ is the $x$ coordinate of $e^s_{y_i}$, $e^s_{y_i}(2)$ is the $y$ coordinate of $e^s_{y_i}$.}
\label{fig: Errors using the resilient controller}
\end{figure}
\begin{figure}[!h]
    \centering
\subfigure[]
{\includegraphics[width=8cm]{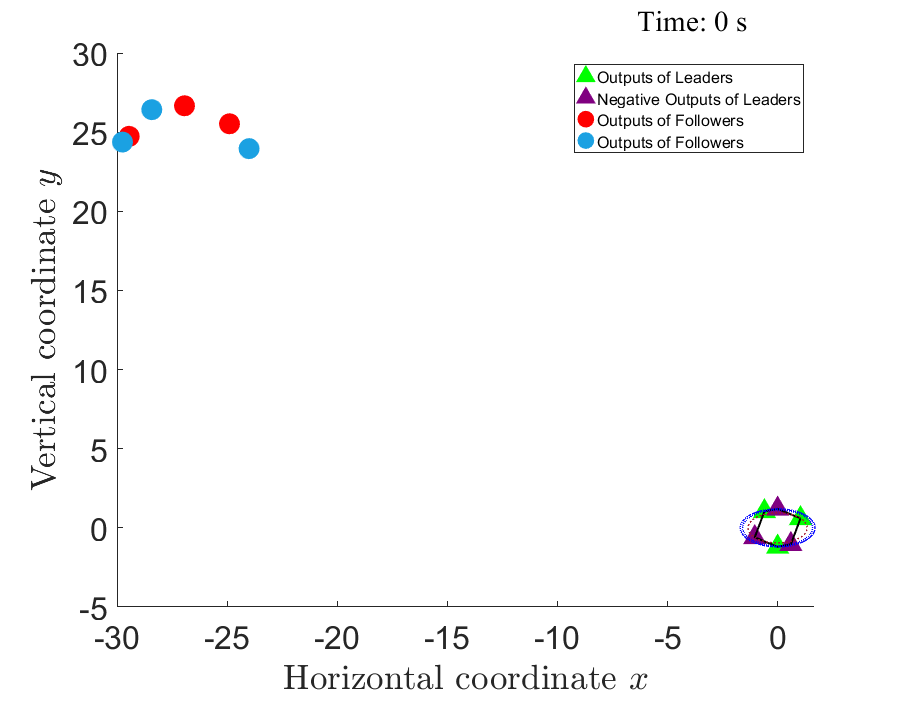}}
    \subfigure[]
{\includegraphics[width=8cm]{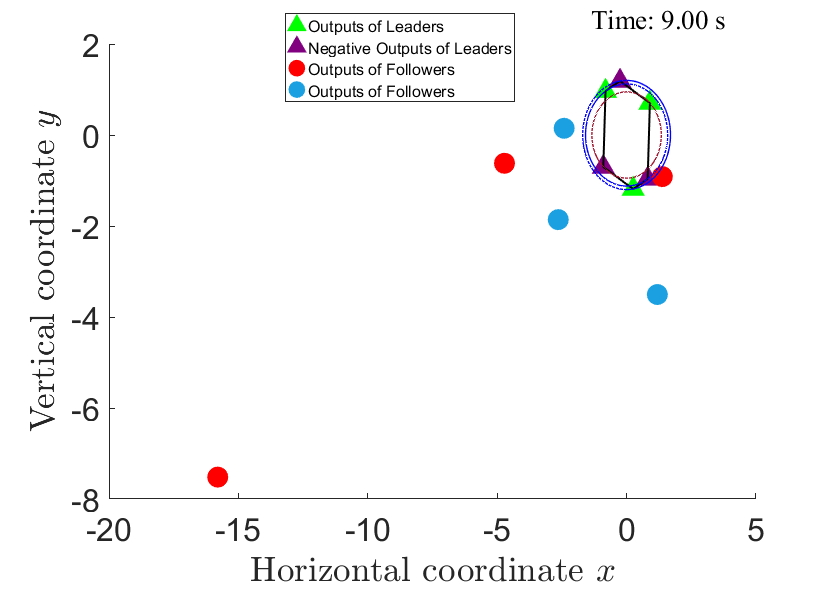}}
\subfigure[]
{\includegraphics[width=8cm]{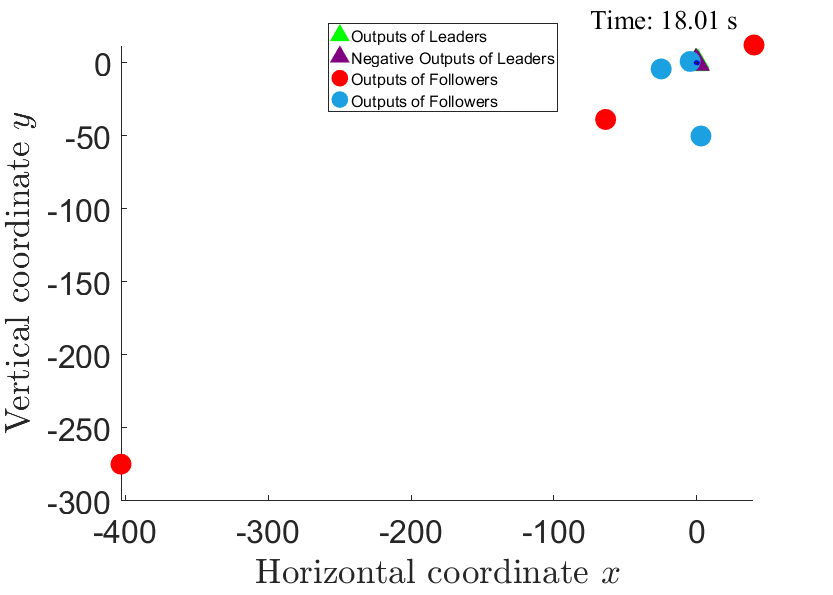}}
    \caption{Leader-follower motion evolution using the standard control protocols: (a) At $0$ s. (b) At $7$ s.(c) At $13$ s.}
    \label{fig: Leader-follower motion evolution without a resilient controller.}
\end{figure}
\begin{figure}[!h]
    \centering
    \subfigure[]    {\includegraphics[width=7.6cm]{0s.png}}
\subfigure[]
{\includegraphics[width=7.6cm]{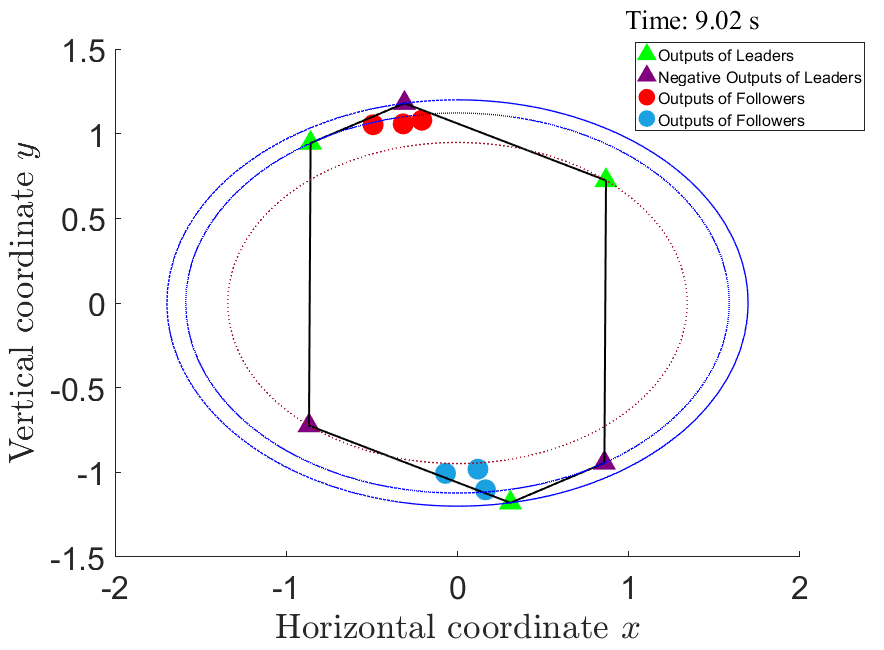}}  
        \subfigure[]
{\includegraphics[width=7.6cm]{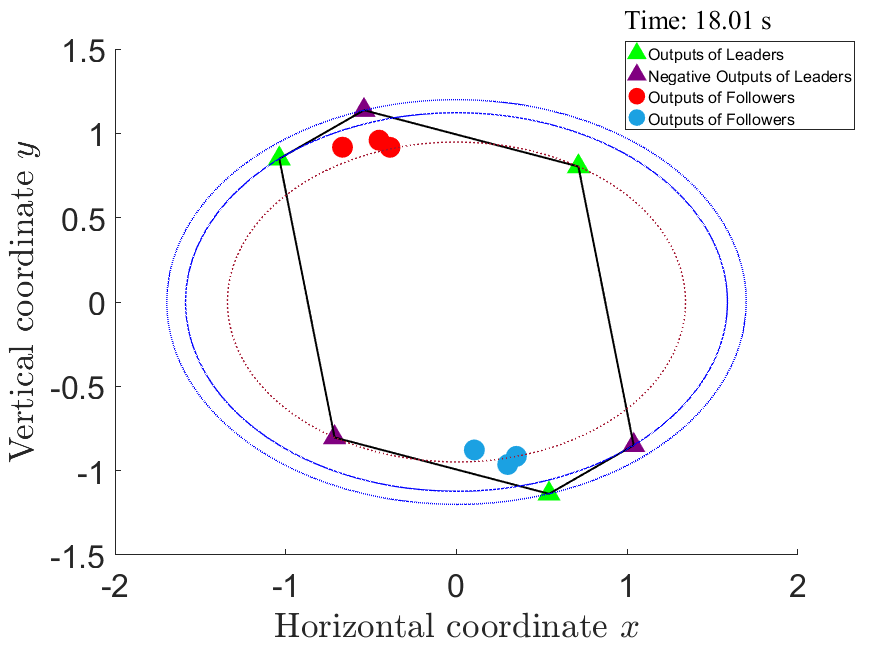}} 
    \captionsetup{justification=centering}
    \caption{Leader-follower motion evolution using the proposed resilient controller: (a) At $0$ s. (b) At $9$ s. (c) At $18$ s.}
    \label{fig: Leader-follower motion evolution with the proposed controller.}
\end{figure}

Based on Lemma~\ref{le: the solution of the PABOC problem}, the bipartite output containment error in \eqref{eq: e^s_y further manipulated} serves to characterize the containment performance of the followers. Fig. \ref{fig: Errors without resilient controller} shows the evolution of the bipartite output containment errors using the standard bipartite containment control protocols described by \eqref{eq: standard ctrl prcls}. As seen, the bipartite output containment errors diverge due to the EU-FDI attacks after $8$ s. Fig.~\ref{fig: Errors using the resilient controller} shows the evolution of the bipartite output containment errors using the proposed resilient control protocols. As seen, after injecting the EU-FDI attacks at $8$ s, $e^s_{y_i}$ stays UUB for each follower, which shows that the UUB convergence performance is achieved under EU-FDI attacks.

Fig.~\ref{fig: Leader-follower motion evolution without a resilient controller.} shows the leader-follower motion evolution using the standard bipartite output containment control protocols. The three hollow circles are the trajectories of the leaders. As shown in Fig.~\ref{fig: Leader-follower motion evolution without a resilient controller.} (b), before the attack initiation at $8$ s, the standard control protocols achieve the bipartite output containment control objective, where the followers converge to the convex hull spanned by the outputs and negative outputs of the 3 leaders. However, as seen in Fig.~\ref{fig: Leader-follower motion evolution without a resilient controller.} (c), the followers' trajectories diverge and fail to achieve the PABOC objective after the initiation of the EU-FDI attacks at $8$ s. Fig.~\ref{fig: Leader-follower motion evolution with the proposed controller.} shows the leader-follower motion evolution using the proposed resilient control protocols. As seen from Fig.~\ref{fig: Leader-follower motion evolution with the proposed controller.} (c), after the initiation of the EU-FDI attacks, the followers remain confined to a small neighborhood around the convex hull spanned by the outputs and negative outputs of the three leaders, which validates the enhanced resilient performance of the proposed cyber-physical defense strategies against EU-FDI attacks on both CPL and OL.

\begin{figure}[!h]
\centering
\includegraphics[width=3in]{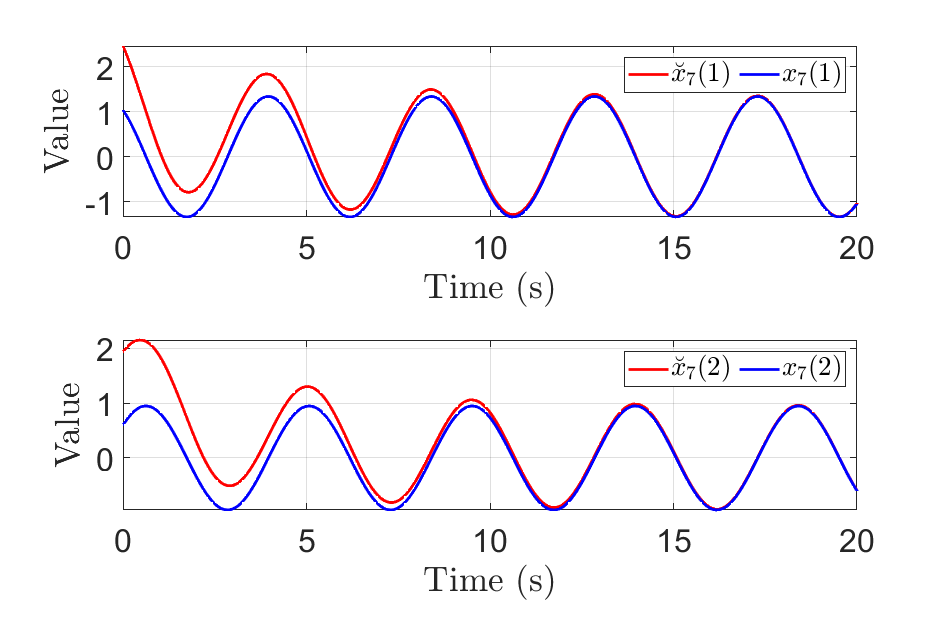}
\captionsetup{justification=centering}
\caption{Comparison of masked and unmasked data of $x_7$.}
\label{fig: Compare_masked_unmasked_leader1_state}
\end{figure}

\begin{figure}[!h]
\centering
\includegraphics[width=3in]{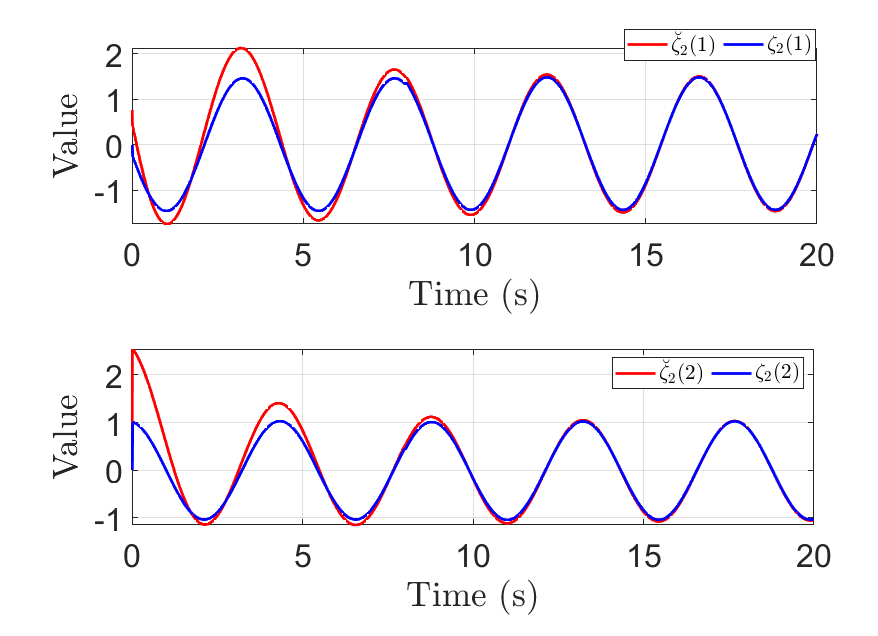}
\captionsetup{justification=centering}
\caption{Comparison of masked and unmasked data of $\zeta_2$.}
\label{fig: Compare_zeta_masked_zeta}
\end{figure}

Fig.~\ref{fig: Compare_masked_unmasked_leader1_state} and \ref{fig: Compare_zeta_masked_zeta} show the comparison of the masked and unmasked data of leader $x_7$ and follower $x_2$, respectively. The transmitted data from leader $x_7$ and follower $x_2$ are required to construct observer state $\zeta_5$. As seen, the transmitted data are masked using the mask function which preserves the initial conditions and the real value of the data. there exist errors between the real data values and the masked data value in the beginning of the time interval, and the error converges to 0 as time progresses. The convergence time can be adjusted by tuning the parameters in \eqref{eq: the mask function for leader} and \eqref{eq: our mask function for followers} appropriately. 

\section{Exprimental validation: Special Case study for power Microgrids}
In this section we have implemented our algorithm to microgrids which is a specific case of signed graph. 
Based on \cite{zuo2020distributed}, that the standard cooperative secondary control for DC microgrids transforms the problem into consensus control for first-order linear MAS, aiming to regulate average voltage to a global reference and ensure proportional load sharing. 
We implemented our proposed observer layer design to estimate the global average voltage under attacks in the OL, as detailed below:
\begin{align*}
\label{eq4}
&\dot{\bar{V}}_i=\dot{V}_i+\exp{(\vartheta_i)} \sum\limits_{j \in {\mathcal{N}_i}} a_{i j}\left(\breve{\bar{V}}_{j}-\bar{V}_i\right) + \gamma^{OL}_i \\
&\dot{\vartheta}_i = q_i\xi_{i_{MG}}^{\top}\xi_{i_{MG}}
\end{align*}
where $\xi_{i_{MG}} = \sum\limits_{j \in {\mathcal{N}_i}} a_{i j}\left(\breve{\bar{V}}_{j}-\bar{V}_i\right)$. To ensure bounded global voltage regulation and proportional load sharing under unknown unbounded FDI attacks, we propose the following attack-resilient secondary control protocols for the microgrid as a special case of signed digraph heterogeneous
MAS:  
\begin{equation*}
\label{eq3}
u_i=\left(g_i\left(V_{\mathrm{ref}}-\bar{V}_i\right)+\sum_{j \in \mathcal{N}_i} a_{i j}\left(R_{j}^{\mathrm{vir}} I_{j}-R_i^{\mathrm{vir}} I_i\right)\right)+ {\gamma^a_i} -\hat \gamma^a_i,
\end{equation*} 

\begin{figure}[!h]
\centering
\includegraphics[width=6cm]{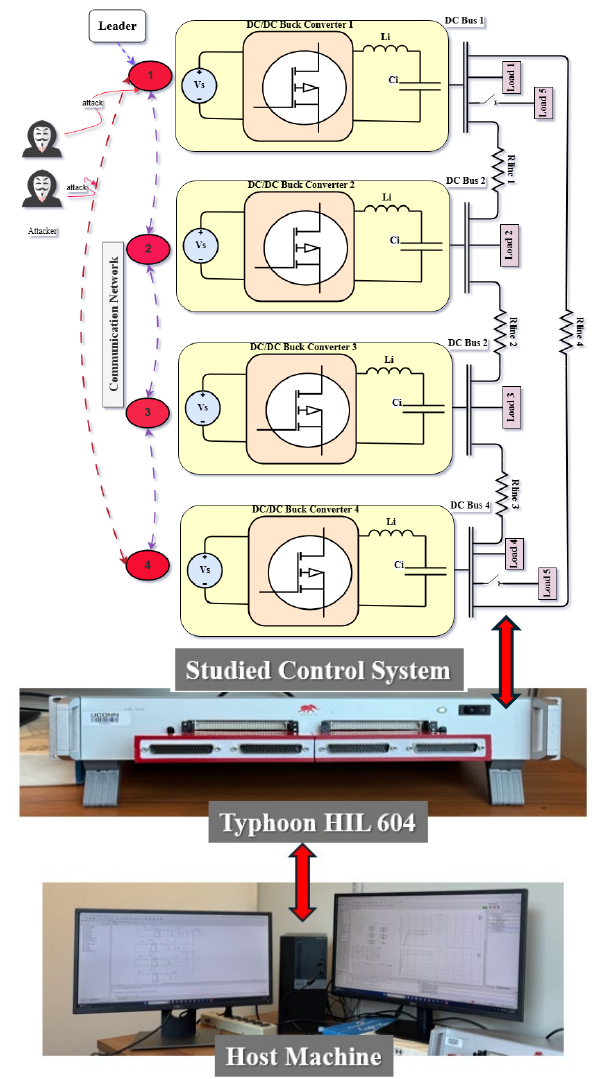}
\captionsetup{justification=centering}
\caption{Microgrid structure.}
\label{fig: MG structure}
\end{figure}

A low-voltage DC microgrid (MG), depicted in Fig.~\ref{fig: MG structure}, is modeled to evaluate the effectiveness of the proposed control methodology. The practical validation of both the control protocol and the DC MG model is conducted using four DC-DC converters emulated on a Typhoon HIL 604 system, as illustrated in Fig.~\ref{fig: MG structure}. Also, the communication network, depicted in this figure, is assumed to have bidirectional links which is a special case study of the proposed method. This setup ensures a high-fidelity representation of real-world operating conditions. Each power source is interfaced through a buck converter. While the converters share similar topologies, they are designed with different current ratings: $I_{1,2,3,4}^{\operatorname{rated}}=(6,3,3,6)$, and virtual impedances: $R_{1,2,3,4}^{\operatorname{vir}}=(2,4,4,2)$. The key parameters of the converters include capacitance $C=2.2\,\mathrm{mF}$, inductance $L=2.64\,\mathrm{mH}$, switching frequency $f_{s}=60\,\mathrm{kHz}$, line resistance $R_{\text{line}}=0.1\,\Omega$, load resistance $R_{L}=10\,\Omega$, reference voltage $V_{\text{ref}}=48\,\mathrm{V}$, and input voltage $V_{\mathrm{in}}=80\,\mathrm{V}$. The rated voltage of the DC MG is maintained at $48\,\mathrm{V}$.

\begin{figure}[!h]
\centering
\includegraphics[width=9cm]{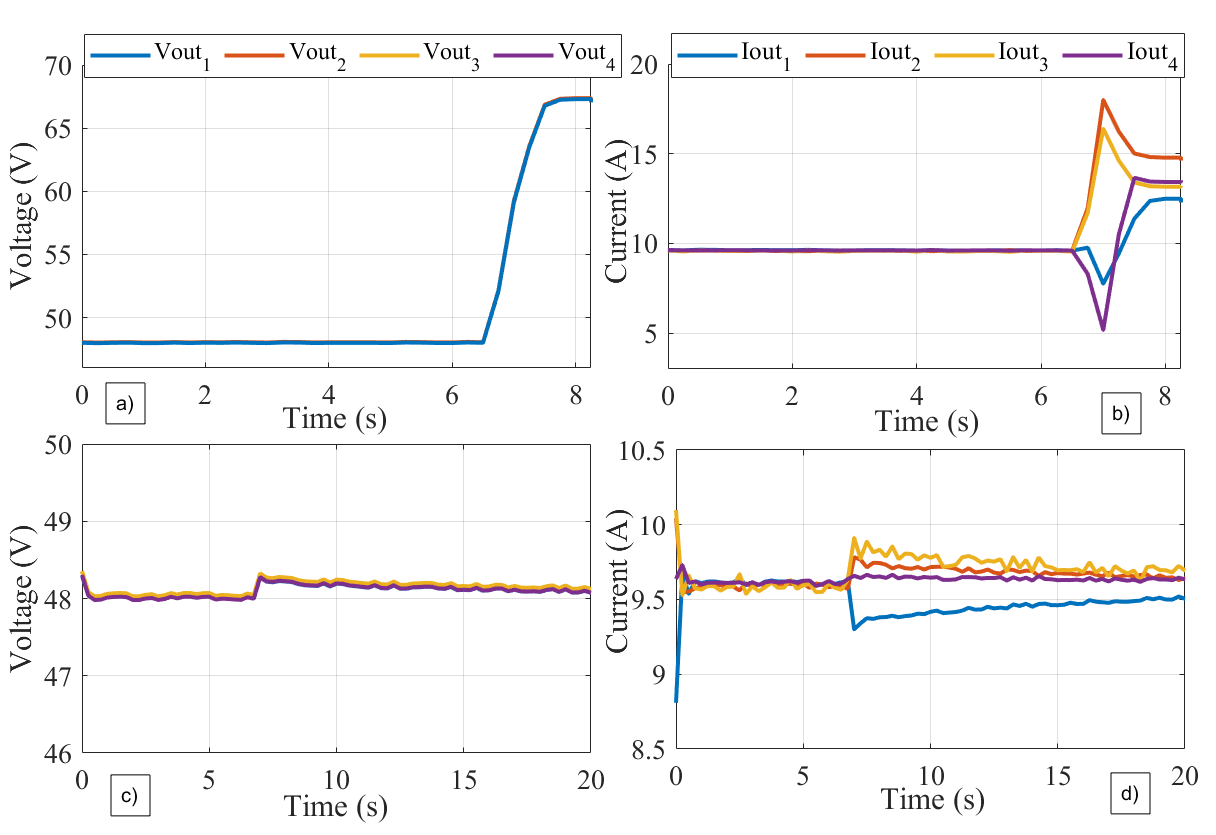}
\captionsetup{justification=centering}
\caption{Performance of the (a) and (b)  Conventional, (c) and (d) proposed attack-resilient control approach
in the case of EU-FDI attacks}
\label{fig: HIL_results}
\end{figure}

To assess the suggested controller performance, a comparison is made with the conventional resilient controller. 
The test lasts from $0$ to $20$ seconds. This part discusses the EU-FDI attack model, which involves injecting EU-FDI attacks at the local control input and observer layer of each converter by selecting $ {\gamma^a_i}=[3\mathrm{exp}\left(0.1t\right)\:4\mathrm{exp}\left(0.2t\right)\:0.5\mathrm{exp}\left(0.2t\right)\:0.1\mathrm{exp}\left(0.3t\right)]^\mathrm{T}$, $ \gamma^{OL}_i=[0.5\mathrm{exp}\left(0.1t\right)\:0.2\mathrm{exp}\left(0.1t\right)\:0.5\mathrm{exp}\left(0.2t\right)\:0.1\mathrm{exp}\left(0.3t\right)]^\mathrm{T}$, $\forall i=1,2,3,4 $.   
Initially, the conventional secondary controller is illustrated to be ineffective when subjected to an EU-FDI attacks on the MG system. Evidently as shown in Fig.~\ref{fig: HIL_results} (a and b), following the onset of the FDI attack at approximately $ t=6.3 s$, both bus voltage and current exhibit a continuous rise, indicating the incapacity of the conventional secondary controller to fulfill control objectives in the presence of such attacks.

However, Fig.~\ref{fig: HIL_results} (c and d) illustrates that the proposed resilient control method ensures the terminal voltages of the converters remain bounded and close to the desired value of $48\,\mathrm{V}$, even under EU-FDI attacks. Additionally, the supplied currents are properly shared despite these attacks. Our attack-resilient protocol maintains system stability and keeps voltages and currents within acceptable operational limits.

\section{Conclusion}
This paper has proposed a fully-distributed privacy-preserving attack-resilient bilayer defense framework to address the PABOC problem for heterogeneous MASs in the face of EU-FDI attacks on both the CPL and OL in the presence of eavesdroppers. First, an attack-resilient dynamic observer utilizing neighborhood relative information exchanged on the OL is designed to estimate convex combinations of the states and negative states of the leaders. To ensure the security of transmitted data, a privacy-preserving mechanism is incorporated into the observer design, masking critical information during communication and enhancing privacy against potential eavesdropping. The observer effectively addresses EU-FDI attacks on the OL, guaranteeing UUB estimation of the leaders' states. Then, using the observer's state, a fully-distributed attack-resilient local controller is developed to address additional EU-FDI attacks on local actuators. Rigorous Lyapunov stability analysis has established the theoretical soundness of the proposed framework, ensuring UUB consensus, stability, and privacy, in the face of adversarial attackers and eavesdroppers. The enhanced resilience of the proposed defense strategies has been validated through comparative simulation case studies on heterogeneous MASs and the application in DC microgrids, demonstrating the effectiveness and practicality of the proposed approach.

\ifCLASSOPTIONcaptionsoff
  \newpage
\fi
\appendix
\noindent\textbf{Proof of Theorem 1.}
\vspace{3mm}
To prove Problem \ref{pro: paboc} is solved, we need to prove (ii) in Problem~\ref{pro: paboc}. The proof of the privacy preservation by \eqref{eq: the mask function for leader} is analogous to that in \cite{altafini2019dynamical} and is omitted here for brevity.

The proof of the privacy preservation by \eqref{eq: our mask function for followers} is as follows.

$C1: h(0,x_i, \frak{p}_i)) = \left(1 + \phi\right)\left( x_i(t) + \wp_i\right)\neq x_i(t)$. Therefore, $C1$ is satisfied.

$C2:$ In \eqref{eq: our mask function for followers}, the knowledge of $\dot{x}_i(t)$ and $\dot h_i(t, x_i, \frak{p}_i)$ is insufficient to uniquely determine the parameters $\frak{p}_i =\{\phi_i, \sigma_i, \wp_i, \vartheta_i\}$ which are private to each agent. The adversaries are unable to reconstruct $x_0$ which requires solving a non-linear system involving unknowns parameters ($\frak{p}_i$), making the task computationally infeasible. Therefore, $h_i(t, x_i, \frak{p}_i)$ adheres to condition $C2$.

$C3:$ The mask function is defined as:
\[
\breve{\zeta}_i(t, x_i) = \left( 1 + \phi_i^f e^{-\sigma_i^f t} \right) \left( \zeta_i + \wp_i^f e^{-\vartheta_i (t)} \right),
\]
where \( \phi_i^f, \sigma_i^f > 0 \) control the exponential decay, \( \zeta_i \) represents the state variable, \( \wp_i^f \) denotes additional data, and \( \vartheta_i > 0 \) is a time-varying signal.

To determine whether the masked function belongs to an \( \epsilon \)-neighborhood of \( x^* \in \mathbb{R}^n \), assume the initial condition satisfies \( \|x_0 - x^*\| < \epsilon \). At \( t = 0 \), the masked function simplifies to:
\[
\breve{\zeta}_i(0, x_i) = (1 + \phi_i^f) (\zeta_i + \wp_i^f),
\]
where \( \zeta_i \) is the state and \( \wp_i^f \) represents additional data. The distance between the masked value and \( x^* \) is then:
\[
\|\breve{\zeta}_i(0, x_i) - x^*\| = \left\| (1 + \phi_i^f) (\zeta_i + \wp_i^f) - x^* \right\|.
\]

Applying the triangle inequality:
\[
\|\breve{\zeta}_i(0, x_i) - x^*\| \leq \| (1 + \phi_i^f) \zeta_i - x^* \| + \| (1 + \phi_i^f) \wp_i^f \|,
\]
where, the term \( (1 + \phi_i^f) \zeta_i \) scales the state \( \zeta_i \), and the factor \( 1 + \phi_i^f > 1 \) generally amplifies the deviation. The term \( \wp_i^f \), representing additional data in the mask function, introduces an offset that contributes further to the overall distance. Thus, while the masked function incorporates the state \( \zeta_i \), the scaling factor \( 1 + \phi_i^f \) and the perturbation \( \wp_i^f \) imply that the resulting value does not, in general, belong to an \( \epsilon \)-neighborhood of \( x^* \). Specifically, for sufficiently small \( \epsilon > 0 \), the presence of these terms can lead to deviations that exceed the original neighborhood.

$C4:$ From \eqref{eq: theta i dot} and the description following it, $\vartheta_i(0) = 0$ and $\dot \vartheta_i(t) > 0$. From the construction of $h(t,x_i, \frak{p}_i)) = \left(1 + \phi^f_ie^{-\sigma^f_it}\right)\left( x_i + \wp^f_i e^{-\vartheta_i(t)} \right)$, it can be readily seen that, $h_i(t, x_i, \frak{p}_i)$ strictly increases in $x_i$ for each fixed $t$ and $\frak{p}_i, \; i = 1, 2, \dots, N$.

$C5:$ \[ |h_i(t, x_i, \frak{p}_i) - x_i| = \wp^f_i e^{-\vartheta_i(t)} + \phi^f_ie^{-\sigma^f_it}x_i + \wp^f_i \phi^f_i e^{-\sigma^f_i t - \vartheta_i(t)}\], $\vartheta_i(0) = 0$ and $\dot \vartheta_i(t) > 0$. By inspection, it is clear that is monotonically decreasing with \(t\) for each fixed $x_i$ and it is straightforward to verify that $\lim_{t \to \infty} h_i(t, x_i, \frak{p}_i) = x_i, \; i = 1, 2, \dots, N$. Therefore, $C5$ is satisfied.

Expanding \eqref{eq: the mask function for leader} yields:
\begin{equation}
h(t, x_r, \mathfrak{p}_i) = x_r + x_r d^l_r + c^l_r,
\label{eq: expanded mask function leader}
\end{equation}
where $d^l_r = (\phi^l_i e^{-\sigma^l_i t})\rightarrow 0$, $c^l_r = (\wp^l_i e^{-\delta^l_i t} + \wp^l_i \phi^l_i e^{-(\sigma^l_i + \delta^l_i)t})\rightarrow 0$. Note that $\lim_{t \to \infty}{c^l_r} = 0$. Based on Assumption~\ref{ass: eig of S},
\( \lim_{t \to \infty}{x_r d^l_r} = 0 \). Denote \( \mathring l_i =  x_r d^l_r + c^l_r\), then $\lim_{t \to \infty}{\mathring l_i} = 0$. Select the $\zeta_i$ as the masked variable and expanding \eqref{eq: our mask function for followers} yields:
\begin{equation}
h(t, \zeta_i, \mathfrak{p}_i) = \zeta_i + \zeta_i b_i^f + c^f_i,
\label{eq: expanded mask function follower}
\end{equation}
where $b_i^f = \phi^f_i e^{-\sigma^f_i t} \rightarrow 0$ and $c^f_i = (\wp^f_i e^{-\vartheta_i(t)} + \wp^f_i \phi^f_i e^{-(\sigma^f_i + \vartheta_i(t))}) \rightarrow 0$. Denote $\mathring f_i = \zeta_i b_i^f + c^f_i$. Plugging \eqref{eq: expanded mask function leader} and \eqref{eq: expanded mask function follower} into \eqref{eq: masked xii} yields
\begin{equation}
\begin{gathered}
\breve\xi_i= \sum_{j \in \mathscr{F}} (a_{ij}\mathring f_j) +
\sum_{j \in \mathscr{F}} (a_{ij}{\zeta}_j- |a_{ij}|\zeta_{i}) + \sum_{r \in \mathscr{L}}(g_{ir} x_{r} \\
- |g_{ir}|\zeta_{i}) + \sum_{r \in \mathscr{L}} (g_{ir}\mathring l_i).\hfill
\end{gathered}
\label{eq: masked xii manipulated}
\end{equation}
Denote $\mathring{z}_i = \sum_{j \in \mathscr{F}} (a_{ij}\mathring f_j) + \sum_{r \in \mathscr{L}} (g_{ir}\mathring l_i)$. It follows that $\breve{\xi}_i = \xi_i + \mathring{z}_i$ and $\breve\xi = \xi + \mathring{z}$, where, $\breve\xi= {[ {\breve\xi_1^{\top},...,\breve\xi_N^{\top}} ]^{\top}}, \xi= {[ {\xi_1^{\top},...,\xi_N^{\top}} ]^{\top}}$ and $\mathring{z}= {[ {\mathring{z}_1^{\top},...,\mathring{z}_N^{\top}} ]^{\top}}$.

From Lemma \ref{le: the solution of the PABOC problem}, to prove that Problem \ref{pro: paboc} is solved, we also need to prove that $e^s_y$ is UUB. Note that $e^s_y$ in \eqref{eq: global form: Neighborhood Bipartite Containment Error} can be written as
\begin{equation}
\begin{gathered}
e^s_y=-\sum_{\nu \in \mathscr{L}} (\Phi_{\nu}^s \otimes I_z) \Bigg( y - \hfill\\
    \sum_{r \in \mathscr{L}} \left( \left( \sum_{k \in \mathscr{L}} \Phi_k^s \right)^{-1} \otimes I_z \right) \left(\left(\frac{1}{M} \bar{\mathcal{L}} + \mathcal{G}_r \right) \otimes I_z\right) \bar{y}_r\Bigg)\\
    =-\sum_{\nu \in \mathscr{L}} (\Phi_{\nu}^s \otimes I_z) \Bigg(\operatorname{diag}(C_i)x -\hfill\\
    \sum_{r \in \mathscr{L}} \left( \left( \sum_{k \in \mathscr{L}} \Phi_k^s \right)^{-1} \otimes I_z \right) \left(\left(\frac{1}{M} \bar{\mathcal{L}} + \mathcal{G}_r \right) \otimes I_z\right) \hfill\\
    \times(I_N\otimes R)\bar x_r\Bigg)\hfill\\
    = -\sum_{\nu \in \mathscr{L}} (\Phi_{\nu}^s \otimes I_z) \Bigg(\operatorname{diag}(C_i)x -(I_N\otimes R)\times\hfill\\
    \sum_{r \in \mathscr{L}} \Bigg( \left( \sum_{k \in \mathscr{L}} \Phi_k^s  \right)^{-1} \left( \frac{1}{M} \bar{\mathcal{L}} + \mathcal{G}_r \right) \otimes I_l \Bigg) \bar x_r\Bigg)\hfill\\
    =-\sum_{\nu \in \mathscr{L}} (\Phi_{\nu}^s \otimes I_z) \Bigg(\operatorname{diag}(C_i)x -\operatorname{diag}(C_i\Pi_i)\hfill\\
    \times\sum_{r \in \mathscr{L}} \Bigg( \left( \sum_{k \in \mathscr{L}} \Phi_k^s  \right)^{-1} \left( \frac{1}{M} \bar{\mathcal{L}} + \mathcal{G}_r \right) \otimes I_l \Bigg) \bar x_r\Bigg)\hfill\\
    =-\sum_{\nu \in \mathscr{L}} (\Phi_{\nu}^s \otimes I_z) \operatorname{diag}(C_i)\Bigg(\varepsilon + \operatorname{diag}(\Pi_i)\hfill\\
    \times\Bigg(\zeta - \sum_{r \in \mathscr{L}} \Bigg( \left( \sum_{k \in \mathscr{L}} \Phi_k^s  \right)^{-1} \left( \frac{1}{M} \bar{\mathcal{L}} + \mathcal{G}_r \right) \otimes I_l \Bigg)\hfill\\ \times \bar x_r\Bigg)\Bigg)\Bigg),\hfill
\end{gathered}
\label{eq26}
\end{equation}
where $\varepsilon = {[ {\varepsilon_1^{\top},...,\varepsilon_N^{\top}} ]^{\top}}$, $\zeta= {[ {\zeta_1^{\top},...,\zeta_N^{\top}} ]^{\top}}$ and $\bar x_r = {[ {x_{N+1}^{\top},...,x_{N+M}^{\top}} ]^{\top}}$. Define the following global compensator containment error
\begin{equation}
\begin{gathered}
\delta=\zeta - \sum_{r \in \mathscr{L}} \Bigg( \left( \sum_{k \in \mathscr{L}} \Phi_k^s  \right)^{-1} \left( \frac{1}{M} \bar{\mathcal{L}} + \mathcal{G}_r \right) \otimes I_l \Bigg) \\
\times\bar x_r.\hfill
\end{gathered}
\label{eq: global delta}
\end{equation}
Then, we obtain
\begin{equation}
e^s_y=-\sum_{\nu \in \mathscr{L}} (\Phi_{\nu}^s \otimes I_z) \operatorname{diag}(C_i)\big(\varepsilon + \operatorname{diag}(\Pi_i)\delta\big).\hfill
\label{eq26}
\end{equation}
To show that $e^s_y$ is UUB, we will prove that $\varepsilon$ and $\delta$ are UUB in the following analysis.

Note that the global form of \eqref{eq: xii} 
\begin{equation}
\begin{gathered}
\xi= -\sum_{\nu \in \mathscr{L}} (\Phi_{\nu}^s \otimes I_l) \Bigg( \zeta - \hfill\\
    \sum_{r \in \mathscr{L}} \left( \left( \sum_{k \in \mathscr{L}} \Phi_k^s \right)^{-1} \otimes I_l \right) \left(\left(\frac{1}{M} \bar{\mathcal{L}} + \mathcal{G}_r \right) \otimes I_l\right) \bar x_r\Bigg)\hfill\\
    = -\sum_{\nu \in \mathscr{L}} (\Phi_{\nu}^s \otimes I_l) \Bigg(\zeta -\hfill\\
    \sum_{r \in \mathscr{L}} \Bigg( \left( \sum_{k \in \mathscr{L}} \Phi_k^s  \right)^{-1} \left( \frac{1}{M} \bar{\mathcal{L}} + \mathcal{G}_r \right) \otimes I_l \Bigg) \bar x_r\Bigg)\hfill\\
    = -\sum_{\nu \in \mathscr{L}} (\Phi_{\nu}^s \otimes I_l)\delta ,\hfill
\end{gathered}
\label{eq: global xi}
\end{equation}
Since $\sum_{\nu \in \mathscr{L}} (\Phi_{\nu}^s \otimes I_l)$ is nonsingular based on Lemma \ref{le: positive real part eighenvalues and non-negative matrices}, to prove that $\delta$ is UUB is equivalent to proving that $\xi$ is UUB.

The global form of $\dot \zeta_i$ in \eqref{eq: zetai dot} is
\begin{equation}
\begin{gathered}
\dot\zeta = (I_N\otimes S)\zeta + \operatorname{diag}(\exp(\vartheta_i))(\xi + \mathring{z}) + \gamma^{OL}.\hfill
\end{gathered}
\label{eq: global form of zeta_i dot}
\end{equation}
where $\gamma^{OL}= {[ {{\gamma^{OL}_1}^{\top},...,{\gamma^{OL}_N}^{\top}} ]^{\top}}$.
Then the time derivative of $\xi$ in \eqref{eq: global xi} is
\begin{equation}
\begin{gathered}
\dot{\xi} = -\sum_{\nu \in \mathscr{L}} (\Phi_{\nu}^s \otimes I_l) \Bigg(\dot\zeta -\hfill\\
    \sum_{r \in \mathscr{L}} \Bigg( \left( \sum_{k \in \mathscr{L}} \Phi_k^s  \right)^{-1} \left( \frac{1}{M} \bar{\mathcal{L}} + \mathcal{G}_r \right) \otimes I_l \Bigg) \dot{\bar x}_r\Bigg)\hfill\\
    = -\sum_{\nu \in \mathscr{L}} (\Phi_{\nu}^s \otimes I_l) \Bigg((I_N\otimes S)\zeta + \big(\operatorname{diag}(\exp(\vartheta_i))\otimes{I_l}\big)(\xi + \mathring{z})\hfill\\
    + \gamma^{OL} - \sum_{r \in \mathscr{L}} \Bigg( \left( \sum_{k \in \mathscr{L}} \Phi_k^s  \right)^{-1} \left( \frac{1}{M} \bar{\mathcal{L}} + \mathcal{G}_r \right) \otimes I_l \Bigg)\hfill\\
    \times(I_N\otimes S)\bar x_r\Bigg)\hfill\\
=(I_N\otimes S)\xi-\sum_{r \in \mathscr{L}} (\Phi_{r}^s \otimes I_l)\big(\operatorname{diag}(\exp(\vartheta_i))\otimes{I_l}\big)(\xi + \mathring{z})\hfill\\
-\sum_{r \in \mathscr{L}} (\Phi_{r}^s \otimes I_l)\gamma^{OL}.\hfill
\end{gathered}
\label{eq: global xi dot}
\end{equation}
We consider the following Lyapunov function candidate
\begin{equation}
V^{'} = \frac{1}{2}\sum\limits_{i = 1}^N \xi_i^{\top}\xi_i \exp(\vartheta_i).
\label{eq: Lyapunov function}
\end{equation}
The time derivative of $V^{'}$ along the trajectory of \eqref{eq: global xi dot} is given by
\begin{equation}
\begin{gathered}
  \dot V^{'} = \sum\limits_{i = 1}^N \big(\xi_i^{\top}{\dot{\xi}_i} \exp(\vartheta_i) + \frac{1}{2}\xi^{\top}_i\xi_i\exp(\vartheta_i)\dot \vartheta_i\big)\hfill \\
  = \xi^{\top}\operatorname{diag}\big(\exp(\vartheta_i)\otimes I_l\big)\dot{\xi} + \frac{1}{2}\xi^{\top}_i\big(\operatorname{diag}(\exp(\vartheta_i)\dot \vartheta_i)\otimes I_l\big)\hfill\\
  \times \xi \hfill\\
  = \xi^{\top}\operatorname{diag}\big(\exp(\vartheta_i)\otimes I_l\big)
  \bigg((I_N\otimes S)\xi - \sum_{r \in \mathscr{L}} (\Phi_{r}^s \otimes I_l)\hfill\\
  \times\big(\operatorname{diag}(\exp(\vartheta_i))\otimes{I_l}\big)(\xi + \mathring{z})
  - \sum_{r \in \mathscr{L}} (\Phi_{r}^s \otimes I_l)\gamma^{OL}\bigg) + \frac{1}{2}\xi^{\top}\hfill\\
  \times\big(\operatorname{diag}(\dot \vartheta_i)\otimes I_l\big)\big(\operatorname{diag}(\exp(\vartheta_i))
  \otimes I_l\big)\xi\hfill\\
  \leqslant \sigma_{\max}(S)\|\big(\operatorname{diag}(\exp(\vartheta_i))\otimes{I_l}\big)\xi\|\|\xi\|- \sigma_{\min}\big(\sum_{r \in \mathscr{L}}\Phi_{r}^s\big)\hfill\\
  \times\|\big(\operatorname{diag}(\exp(\vartheta_i))\otimes{I_l}\big)\xi\|^2 + \sigma_{\min}\big(\sum_{r \in \mathscr{L}}\Phi_{r}^s\big)\hfill\\
  \times\|\big(\operatorname{diag}(\exp(\vartheta_i))\otimes{I_l}\big)\xi\|\|\big(\operatorname{diag}(\exp(\vartheta_i))\otimes{I_l}\big)\mathring{z}\|\hfill\\
  + \sigma_{\max}\big(\sum_{r \in \mathscr{L}}\Phi_{r}^s\big)\|\big(\operatorname{diag}(\exp(\vartheta_i))\otimes{I_l}\big)\xi\|\|\gamma^{OL}\| \hfill\\
  + \frac{1}{2}\max_i(\dot \vartheta_i)\|\big(\operatorname{diag}(\exp(\vartheta_i))\otimes{I_l}\big)\xi\|\|\xi\|\hfill\\
  = - \sigma_{\min}\big(\sum_{r \in \mathscr{L}}\Phi_{r}^s\big)\|\big(\operatorname{diag}(\exp(\vartheta_i))\otimes{I_l}\big)\xi\|\hfill\\
  \times\bigg(\|\big(\operatorname{diag}(\exp(\vartheta_i))\otimes{I_l}\big)\xi\| - \sigma_{\max}(S)\hfill\\
  /\sigma_{\min}\big(\sum_{r \in \mathscr{L}}\Phi_{r}^s\big)\|\xi\| - \|\big(\operatorname{diag}(\exp(\vartheta_i))\otimes{I_l}\big)\mathring{z}\|\hfill\\
  - \sigma_{\max}\big(\sum_{r \in \mathscr{L}}\Phi_{r}^s\big)/\sigma_{\min}\big(\sum_{r \in \mathscr{L}}\Phi_{r}^s\big)\|\gamma^{OL}\| - \frac{1}{2}\max_i(\dot \vartheta_i)\hfill\\
  /\sigma_{\min}\big(\sum_{r \in \mathscr{L}}\Phi_{r}^s\big)\|\xi\|\bigg).\hfill\\
\end{gathered}
\label{eq: time derivative of the Lyapunov function}
\end{equation}
For convenience, denote $\phi_a = \sigma_{\max}(S)
  /\sigma_{\min}\big(\sum_{r \in \mathscr{L}}\Phi_{r}^s\big)$ and $\phi_b = \sigma_{\max}\big(\sum_{r \in \mathscr{L}}\Phi_{r}^s\big)/\sigma_{\min}\big(\sum_{r \in \mathscr{L}}\Phi_{r}^s\big)$, which are both positive constants. To let $\dot V^{'} \leqslant 0$, we need
\begin{equation}
\begin{gathered}
\|\big(\operatorname{diag}(\exp(\vartheta_i))\otimes{I_l}\big)\xi\| - \phi_a\|\xi\|  - \|\big(\operatorname{diag}(\exp(\vartheta_i))\otimes{I_l}\big)\mathring{z}\|\hfill\\
- \phi_b\|\gamma^{OL}\| - \frac{1}{2}\max_i(\dot \vartheta_i)/\sigma_{\min}\big(\sum_{r \in \mathscr{L}}\Phi_{r}^s\big)\|\xi\|\geqslant 0. \hfill
\end{gathered}
\label{eq: critical part of dot V'}
\end{equation}

A sufficient condition to guarantee \eqref{eq: critical part of dot V'} is 
\begin{equation}
\begin{gathered}
\big(\exp(\vartheta_i) - \phi_a - \frac{1}{2}\max_i(\dot \vartheta_i)/\sigma_{\min}\big(\sum_{r \in \mathscr{L}}\Phi_{r}^s\big)\big)\|\xi_i\| - \exp(\vartheta_i)\hfill\\
\times\|\mathring{z}_i\|\geqslant\phi_b\|\gamma^{OL}_i\|.\hfill
\end{gathered}
\label{eq: the sufficient condition of time derivative of the Lyapunov function ge to 0}
\end{equation}

For convenience, we denote $\|\mathring{z}_i\| = \exp(-p_it) \|\xi_i\|$. A sufficient condition to guarantee \eqref{eq: the sufficient condition of time derivative of the Lyapunov function ge to 0} is $\|\xi_i\|\geqslant\phi_b$ and $\exp(\vartheta_i) - \exp(\vartheta_i)\exp(-p_it) - \phi_a- 1/2\max_i(\dot \vartheta_i)/\sigma_{\min}(\sum_{r \in \mathscr{L}}\Phi_{r}^s)\geqslant\|\gamma^{OL}_i\|$. Based on Assumption \ref{ass: attacks}, there exists a positive constant $\kappa^{OL}_i$ such that $\|\gamma^{OL}_i(t)\| \leqslant \exp(\kappa^{OL}_it)$. To prove that $\exp(\vartheta_i) - \exp(\vartheta_i)\exp(-p_it) - \phi_a - 1/2\max_i(\dot \vartheta_i)/\sigma_{\min}(\sum_{r \in \mathscr{L}}\Phi_{r}^s)\geqslant\|\gamma^{OL}_i\|$, we need to prove that $\exp(\vartheta_i) - \exp(\vartheta_i)\exp(-p_it) - \phi_a - 1/2\max_i(\dot \vartheta_i)/\sigma_{\min}(\sum_{r \in \mathscr{L}}\Phi_{r}^s)\geqslant\exp(\kappa^{OL}_it)$. Based on \eqref{eq: theta i dot} and \eqref{eq: masked xii}, when $\|\xi_i\|>\max\{\sqrt{\kappa^{OL}_i/q_i},\phi_b\}$, which guarantees the exponential growth of $\exp(\vartheta_i)$ dominates all other terms, $\exists t_1$ such that $\forall t > t_1$, $\exp(\vartheta_i) - \phi_a - 1/2\max_i(\dot \vartheta_i)/\sigma_{\min}(\sum_{r \in \mathscr{L}}\Phi_{r}^s)\geqslant\exp(\kappa^{OL}_it)$. Hence, we obtain $\forall t > t_1$,

% The sufficient condition of $\exp(\vartheta_i) - \phi_a - 1/2\max_i(\dot \vartheta_i)/\sigma_{\min}(\sum_{r \in \mathscr{L}}\Phi_{r}^s)\geqslant\exp(\kappa^{OL}_it)$ is $\|\xi_i\|\geqslant \sqrt{\kappa^{OL}_i/q_i}$. Hence, when $\|\xi_i\|\geqslant\max\{\phi_b, \sqrt{\kappa^{OL}_i/q_i}\}$,

\begin{equation}
\dot V^{'} \leqslant 0,\:\forall \|\xi_i\|>\max\{\sqrt{\kappa^{OL}_i/q_i},\phi_b\}.
\label{eq: dot Lyapunov function candidate le 0}
\end{equation}
By LaSalle’s
invariance principle \cite{krstic1995nonlinear}, $\xi_i$ is UUB.

Next, we prove that $\varepsilon$ is UUB. 
From \eqref{eq: follower dynamics}, \eqref{eq: output solution}, \eqref{eq: zetai dot}, \eqref{eq: controller ui} and \eqref{eq: Hi}, we obtain the time derivative of \eqref{eq: epsiloni} as
\begin{equation}
\begin{gathered}
  {{\dot \varepsilon }_i} = {{\dot x}_i} - \Pi_i(\dot\zeta_i + \exp(\vartheta_i)\mathring{z}_i) \hfill \\
   = {A_i}{x_i} + {B_i}{K_i}{x_i} + {B_i}{H_i}{\zeta _i} - {B_i}{\hat \gamma^a_i} \hfill \\
   + {B_i}{\gamma^a_i} - {\Pi _i}S{\zeta _i} - {\Pi _i}\exp(\vartheta_i){\xi_i} - \Pi_i\gamma^{OL}_i - \Pi_i\exp(\vartheta_i)\mathring{z}_i \hfill \\
   = \left( {{A_i} + {B_i}{K_i}} \right){\varepsilon _i} + {B_i}{\gamma^a_i} - {B_i}{\hat \gamma^a_i} - {\Pi _i}\exp(\vartheta_i){\xi_i} - \Pi\gamma^{OL}_i\hfill\\
- \Pi_i\exp(\vartheta_i)\mathring{z}_i.\hfill
\end{gathered}
\label{eq: epsilon dot}
\end{equation}
From the above proof, we confirmed ${\xi_i}$ is UUB. Considering Assumption \ref{ass: eig of S}, Based on Assumption \ref{ass: attacks}, there exists a positive constant $\kappa^{OL}_i$ such that \eqref{eq: global xi} and \eqref{eq: global form of zeta_i dot}, we obtain that $\beta_i\equiv{\Pi _i}\exp(\vartheta_i){\xi_i} + \Pi_i\gamma^{OL}_i + \Pi_i\exp(\vartheta_i)\mathring{z}_i$ is bounded. Let ${{\bar A}_i} = {A_i} + {B_i}{K_i}$ and ${\bar Q_i}={Q_i} + K_i^{\top}{U_i}{K_i}$. Note that ${\bar Q_i}$ is positive-definite. From \eqref{eq: solve for P}, $P_i$ is symmetric positive-definite. Consider the following Lyapunov function candidate
\begin{equation}
V_i = \varepsilon_i^TP_i\varepsilon_i,
\label{eq34}
\end{equation}
and its time derivative is given by
\begin{equation}
\begin{gathered}
\dot V_i = 2 \varepsilon_i^TP_i\left(\bar{A}_i \varepsilon_i+B_i \gamma^a_i-B_i \hat\gamma^a_i - \beta_i\right)\hfill\\
\leqslant-\sigma_{\min }\left(\bar{Q}_i\right)\left\|\varepsilon_i\right\|^2+2\left(\varepsilon_i^TP_i B_i \gamma^a_i-\varepsilon_i^TP_i B_i \hat\gamma^a_i\right)\hfill\\
- 2 \varepsilon_i^TP_i\beta_i\hfill\\
\leqslant-\sigma_{\min }\left(\bar{Q}_i\right)\left\|\varepsilon_i\right\|^2+2\left(\varepsilon_i^TP_i B_i \gamma^a_i-\varepsilon_i^TP_i B_i \hat\gamma^a_i\right)\hfill\\
+2 \sigma_{\max }\left(P_i\right)\left\|\varepsilon_i\right\|\left\|\beta_i\right\|.\hfill\\
\end{gathered}
\label{eq35}
\end{equation}

Using \eqref{eq: hat gamma bi} to obtain
\begin{equation}
\begin{gathered}
\varepsilon_i^{\top} P_i B_i \gamma^a_i-\varepsilon_i^{\top} P_i B_i \hat\gamma^a_i \hfill\\
= \varepsilon_i^{\top} P_i B_i \gamma^a_i-\frac{\left\|\varepsilon_i^{\top} P_i B_i\right\|^2}{\left\|\varepsilon_i^{\top} P_i B_i\right\|+\exp \left(-c_i t^2\right)} \exp \left(\hat{\rho}_i\right)\hfill\\
\leqslant \left\|\varepsilon_i^{\top} P_i B_i\right\|\left\|\gamma^a_i\right\|-\frac{\left\|\varepsilon_i^{\top} P_i B_i\right\|^2}{\left\|\varepsilon_i^{\top} P_i B_i\right\|+\exp \left(-c_i t^2\right)} \exp \left(\hat{\rho}_i\right)\hfill\\
=\left\|\varepsilon_i^{\top} P_i B_i\right\|\big(\left\|\varepsilon_i^{\top} P_i B_i\right\|\left\|\gamma^a_i\right\|+\exp(-c_i t^2)\left\|\gamma^a_i\right\|\hfill\\  
-\left\|\varepsilon_i^{\top} P_i B_i\right\|\exp\left(\hat{\rho}_i\right)\big)/\big(\left\|\varepsilon_i^{\top} P_i B_i\right\|+\exp(-c_i t^2)\big).\hfill
\end{gathered}
\label{eq36}
\end{equation}
To prove that $\varepsilon_i^{\top} P_i B_i \gamma^a_i-\varepsilon_i^{\top} P_i B_i \hat\gamma^a_i \leqslant 0$, we need to prove that $\left\|\varepsilon_i^{\top} P_i B_i\right\|\left\|\gamma^a_i\right\| + \exp(-c_i t^2)\left\|\gamma^a_i\right\| -\left\|\varepsilon_i^{\top} P_i B_i\right\|\exp\left(\hat{\rho}_i\right)\leqslant 0$. Define $\upsilon_i = \kappa^a_i/\sigma_{\operatorname{min}}(P_iB_i)$, $\omega_i = 2 \sigma_{\max }\left(P_i\right)\left\|\beta_i\right\|/\sigma_{\min }\left(\bar{Q}_i\right)$. Then, define the compact sets $\Upsilon_i\equiv\ \{\|\varepsilon_i\|\leqslant\upsilon_i\}$ and $\Omega_i\equiv\{\|\varepsilon_i\|\leqslant\omega_i\}$. Considering Assumption \ref{ass: attacks}, there exists a positive constant $\kappa^a_i$ such that $\|\gamma^a_i(t)\| \leqslant \exp(\kappa^a_i t)$. We obtain that $\exp(-c_i t^2)\left\|\gamma^a_i\right\|\rightarrow 0$. Hence, outside the compact set $\Upsilon_i\equiv\ \{\|\varepsilon_i\|\leqslant\upsilon_i\}$, $\exists t_2$, such that $\varepsilon_i^{\top} P_i B_i \gamma^a_i-\varepsilon_i^{\top} P_i B_i \hat\gamma^a_i \leqslant 0$, $\forall t \geqslant t_2$; outside the compact set $\Omega_i\equiv\{\|\varepsilon_i\|\leqslant\omega_i\}$, $-\sigma_{\min }\left(\bar{Q}_i\right)\left\|\varepsilon_i\right\|^2+2 \sigma_{\max }\left(P_i\right)\left\|\varepsilon_i\right\|\left\|\beta_i\right\| \leqslant 0$. Therefore, combining \eqref{eq35}, \eqref{eq36} and \eqref{eq: epsilon dot}, we obtain, outside the compact set $\Upsilon_i\cup\Omega_i$, $\forall t \geqslant t_2$,
\begin{equation}
\dot V_i\leqslant 0.
\label{eq: Vi dot le 0}   
\end{equation}
Hence, by the LaSalle’s
invariance principle, $\varepsilon_i$ is UUB. Consequently, we conclude that $e^s_y$ is UUB. This completes the proof.$\hfill\blacksquare$

\bibliographystyle{IEEEtran}
\bibliography{References}

\end{document}